\documentclass{aa}
\usepackage[utf8]{inputenc}
\usepackage[varg]{txfonts}
\bibpunct{(}{)}{;}{a}{}{,}
\usepackage{graphicx}
\usepackage{placeins}

\newcommand{\disperse}{DisPerSE}

\title{The cosmic web around the Coma cluster from constrained cosmological simulations: I. Filaments connected to Coma at $z = 0$}
\author{Nicola Malavasi\inst{1,3}
\and
Jenny G. Sorce\inst{2,3,4}
\and
Klaus Dolag\inst{1}
\and
Nabila Aghanim\inst{3}
}

\institute{Universit{\"a}ts-Sternwarte, Fakult{\"a}t f{\"u}r Physik, Ludwig-Maximilians-Universit{\"a}t, Scheinerstr. 1, 81679 M{\"u}nchen, Germany
\and
Univ. Lille, CNRS, Centrale Lille, UMR 9189 CRIStAL, F-59000 Lille, France
\and
Universit{\'e} Paris-Saclay, CNRS, Institut d'astrophysique spatiale, 91405, Orsay, France
\and
Leibniz-Institut f\"{u}r Astrophysik (AIP), An der Sternwarte 16, D-14482 Potsdam, Germany}

\date{Received: 23rd December 2022 / Accepted: 22nd May 2023}

\abstract{Galaxy clusters in the Universe occupy the important position of nodes of the cosmic web. They are connected among them by filaments, elongated structures composed of dark matter, galaxies, and gas. The connection of galaxy clusters to filaments is important, as it is related to the process of matter accretion onto the former. For this reason, investigating the connections to the cosmic web of massive clusters, especially well known ones for which a lot of information is available, is a hot topic in astrophysics. In a previous work we performed an analysis of the filament connections of the Coma cluster of galaxies, as detected from the observed galaxy distribution. In this work we resort to a numerical simulation whose initial conditions are constrained to reproduce the Local Universe, including the region of the Coma cluster to interpret our observations in an evolutionary context. We detect the filaments connected to the simulated Coma cluster and perform an accurate comparison with the cosmic web configuration we detected in observations. We perform an analysis of the halos' spatial and velocity distributions close to the filaments in the cluster outskirts. We conclude that, although not significantly larger than the average, the flux of accreting matter on the simulated Coma cluster is significantly more collimated close to the filaments with respect to the general isotropic accretion flux. This paper is the first example of such a result and the first installment in a series of publications which will explore the build-up of the Coma cluster system in connection to the filaments of the cosmic web as a function of redshift.}

\titlerunning{The LSS around Coma from constrained simulations}
\authorrunning{Malavasi et al.}


\begin{document}

\maketitle

\section{Introduction}
The cosmic web \citep{deLapparent1986, Bond1996} is a network of structures present throughout the visible Universe. It is formed by the gravitational collapse of matter, which starts in the primordial Universe from tiny density perturbations \citep{Zeldovich1970a, Zeldovich1970b}. In this process, matter departs from underdense regions, which will become the voids in the final matter distribution. Voids are surrounded by walls (two-dimensional structures). At the intersection of walls are found filaments (one-dimensional, elongated structures). Matter flows inside walls at first and then in filaments (this is especially true at high redshifts) before finally reaching clusters. Galaxy clusters are located at the nodes of the cosmic web and are its most-studied and best-understood components.

The advent of large scale numerical simulations, both N-body \citep{Springel2005} and hydrodynamical, such as Illustris \citep{Vogelsberger2014}, IllustrisTNG \citep{Naiman2018, Marinacci2018, Springel2018, Nelson2018, Pillepich2018}, MAGNETICUM \citep{Hirschmann2014, Dolag2015}, EAGLE \citep{Schaye2015}, and HorizonAGN \citep{Dubois2014}, has allowed us to start investigating the properties of filaments from the theoretical side and to match observational results. Indeed, the filaments of the cosmic web and their impact on galaxies have been extensively investigated in numerical simulations \citep[see e.g.][]{AragonCalvo2010, Cautun2014, Laigle2015, Laigle2018, GVeenaI, GVeenaII, AragonCalvo2019, Kraljic2019, Galarraga2020, Kraljic2020simba, GVeenaIII, Galarraga2021, Gouin2021, Malavasi2022, Galarraga2022}. Results range from the study of the shape, length, and dynamics of filaments, to the matter and volume repartition among the various cosmic web structures, to the connection of the filaments to the clusters and their impact on galaxy properties such as mass, star formation activity, and direction of their angular momentum vector. These studies have found counterparts in similar ones coming from observations \citep{Brouwer2016, Kuutma2017, Malavasi2017, Laigle2018, Kraljic2018, Vulcani2019, Chen2019, Krolewski2019, Malavasi2020Catalogue, Tanimura2020SZ, Tanimura2020X, Bonjean2020, Rost2020, Welker2020}.

Among numerical simulations, constrained ones (such as the CLUES project, \citealt{Gottloeber2010}, and its most recent development: CLONES, \citealt{Sorce2015, Sorce2016vel, Sorce2018, Sorce2021}, used e.g. in HESTIA, \citealt{Libeskind2020}) are gathering increasing importance. Although limited to the local Universe for the moment (at least when peculiar velocities are used as constraints; other techniques which involve the use of densities and whose description is outside the scope of this introduction allow to reach much higher redshifts, see e.g. \citealt{Ata2021}), this kind of simulations is based on real observations of galaxies and allows to reproduce in a realistic way existing structures in a cosmic volume. They have proven increasingly important to interpret the observed properties of clusters in the cosmic web environment, including their evolution, contributing to unveil information normally unavailable through the analysis of clusters found in non-constrained simulations \citep{Sorce2020}.

An important hot topic which has developed following recent advancements in the study of the cosmic web is the connection of the filaments to the galaxy clusters. Indeed filaments have been detected around clusters, both from the gas phase and the galaxy distribution. This ranges from the detection of the tips of the filaments connecting to the cluster in the X-rays \citep{Eckert2015}, to the detection of bridges of matter connecting pairs of close clusters \citep{Akamatsu2017, Bonjean2018, Govoni2019, Reiprich2021, Biffi2022} to a systematic detection of filaments connected to clusters in large-scale simulations \citep{Kuchner2020, Kuchner2021, Gouin2021}, galaxy surveys \citep{Sarron2019, DarraghFord2019, Salerno2019, Salerno2020}, and individual cluster observations \citep{Malavasi2020Coma, Einasto2020, Castignani2022}.

It has been demonstrated that the number of filaments connected to a cluster has an impact on the cluster mass accretion process. More connected clusters are more massive \citep{Codis2018, Sarron2019, DarraghFord2019, Gouin2021} and in a more disturbed dynamical state \citep{Gouin2021, Gouin2022}. Moreover, galaxy evolution in connected structures proceeded faster, as more connected galaxies (which we can consider to be the central galaxies of more connected clusters) are redder, less star-forming and/or more massive \citep{DarraghFord2019, Kraljic2020connectivity}. Moreover, filaments can accrete matter across the virial shock of halos, bringing cold gas that fuels star-formation directly on the forming halo (this is particularly true at high redshift, \citealt{Dekel2009}, \citealt{BennettSijacki2020}).

Among the structures for which the connection to the cosmic web has been investigated in most detail is the Coma cluster of galaxies. The Coma cluster has been studied for several decades \citep{Hubble1931, Biviano1998}, with extensive data sets in all wavelengths, from the X-rays \citep{Briel1992, Neumann2001, Neumann2003} to the SZ signal \citep{Planck2013Coma}, radio \citep{BrownRudnick2011}, and optical \citep{denBrok2011, Adami2005}. The connections of the Coma cluster to other structures of the surrounding cosmic web are already evident from the galaxy distribution and the position of the surrounding clusters \citep[e.g. A1367,][]{West1998, Mahajan2018}. In \citet{Malavasi2020Coma} we explored the connections of the Coma cluster to the filaments of the cosmic web. Filaments \citep{Malavasi2020Catalogue} were identified in the Sloan Digital Sky Survey \citep[SDSS,][]{Abazajian2009, Strauss2002} with the Discrete Persistent Structure Extractor \citep[DisPerSE,][]{Sousbie2011a, Sousbie2011b}. We identified three filaments connected to the cluster, two of which are consistent with the position of known features of the cluster itself. In particular, we identify a filament connecting to Coma on the west side of the cluster (as projected on the plane of the sky) whose position is broadly consistent with both the location of a shock feature visible in the X-ray map of the cluster \citep{Neumann2003, Lyskova2019} as well as with a low-turbulence region visible in the entropy map of the cluster \citep{MirakhorWalker2020}. Another secure filament is identified instead on the north-east side of the cluster (as projected on the plane of the sky), whose position is broadly consistent with the supposed direction of infall of the galaxy NGC4839 and its associated group \citep{Lyskova2019} and with a region of high gas velocity and low temperature and metallicity detected in \emph{XMM-Newton} X-ray maps \citep{Sanders2020}.

The analysis performed in \citet{Malavasi2020Coma} and the supporting evidence exposed in the cited works all point towards the possibility that accretion of matter on the Coma cluster has happened through the filaments and that it is ongoing. As observations can only offer a snapshot of the Coma cluster dynamics fixed in time, we decided to resort to numerical simulations to investigate the accretion of matter on the cluster coming from the filaments. The goal of this paper series is to provide further insight on the evolution and accretion of matter from the filaments onto the Coma cluster. In this first installment, we match the simulated cosmic web to the observed one and we explore the spatial and velocity distributions of matter around the filaments and the simulated Coma cluster at $z = 0$\footnote{We stress that performing the analysis we report in this paper at $z = 0.023$ (the redshift of the Coma cluster) instead of $z = 0$ has no impact on our results whatsoever.}.

This paper is structured as follows: we summarize the observational results of \citet{Malavasi2020Coma} in Section \ref{obsdata} and we introduce the constrained cosmological simulation which we used to reproduce the Coma system in section \ref{numsim}. We describe the algorithm we used to detect the filaments in Section \ref{dispdesc} and we extensively compare the observed and simulated cosmic web around Coma in Section \ref{cwcomparison}. We then discuss the connectivity of the simulated Coma cluster in Section \ref{simcomaconn} and we analyze the dynamics of matter around the cluster and its connected filaments in Section \ref{accretion}. We discuss our results and draw our conclusions in Sections \ref{discussion} and \ref{conclusion}, respectively.

Throughout this paper we use a \citet{Planck2013ParamsSims} cosmology with $H_{0} = 100 \cdot h = 67.8\: \mathrm{km} \cdot \mathrm{Mpc}^{-1} \mathrm{s}^{-1}$, $\Omega_{m} = 0.307$, and $\Omega_{\Lambda} = 0.693$ to be consistent with \citet{Sorce2018} and \citet{Sorce2023velwaves}. We note that this is different from the cosmology used in \citet{Malavasi2020Coma}. We explicit in the text whenever a different cosmology from \citet{Planck2013ParamsSims} is used and we convert between the two whenever needed. Lengths will be provided in units of Mpc/$h$. The only exceptions are for lengths which explicitly refer to values provided in \citet{Malavasi2020Coma}, e.g. the $\pm 75$ Mpc radius used to explore the cosmic web around Coma in that work. In this case, we will provide the corresponding value in Mpc/$h$ in parentheses for reference.

\section{Data and simulations}
In this section we describe our constrained cosmological simulation, which we use to reproduce the Coma cluster and analyze the physical properties of its surrounding LSS, and the observational data with which we compare our simulation. As this work heavily relies on what previously done in \citet{Malavasi2020Catalogue, Malavasi2020Coma}, our observational data essentially amount to what has been used in those works. We recapitulate and summarize the essential information in Section \ref{obsdata}, where we also introduce further data sets we used for specific purposes in this analysis.

\subsection{Observational data}
\label{obsdata}
The analysis by \citet{Malavasi2020Coma} identified the filaments connected to the Coma cluster among those present in the skeleton reconstructed in the Sloan Digital Sky Survey \citep{Abazajian2009} with the Discrete Persistent Structure Extractor (\disperse, \citealt{Sousbie2011a, Sousbie2011b}) and presented in \citet{Malavasi2020Catalogue}. This work used a selection of 566\,452 galaxies from the SDSS DR7 Main Galaxy Sample \citep[MGS,][]{Strauss2002} which have Petrosian r-band magnitude $r_{\mathcal{P}} \leq 17.77$, r-band half-light surface brightness $\mu_{50} \leq 24.5\: \mathrm{mag}\: \mathrm{arcsec}^{-2}$, secure spectroscopic redshift (\textsc{zwarning} = 0, \textsc{zconffinal} > 0.35, and \textsc{zfinal} > 0), and located in a contiguous region in the northern hemisphere. We refer the reader to \citet{Malavasi2020Catalogue, Malavasi2020Coma}, and \citet{Strauss2002} for further details. For the rest of this paper we refer to this selection of objects as the Legacy MGS. In this work, we have enhanced this data set with a measurement of the galaxy masses and K-band magnitudes. In the following, we mainly make use of K-band apparent magnitudes, which we compare with the same quantity available for the halos of our numerical simulation.  We use this comparison to inform our selection of a galaxy sample from our halo population. As for galaxy masses, they are derived as shown in Appendix \ref{appendix_masses} and used only as a further confirmation of our simulated galaxy selection from our halo population and not to derive physical conclusions.

K-band apparent magnitudes for SDSS galaxies were obtained from the New York University Value-Added Galaxy Catalogue \citep[NYU-VAGC,][]{Blanton2005, AdelmanMcCarthy2008, Padmanabhan2008}. This catalogue provides K-band magnitudes for SDSS galaxies from the 2-Micron All Sky Survey \citep[2MASS,][]{Skrutskie2006}. Starting from an initial sample of $2\,506\,754$ galaxies, we select those belonging to the MGS (\textsc{primtarget} \& 64, \textsc{vagc\_select} \& 4). The selection of galaxies in the NYU-VAGC adopted the same criteria as the MGS of \citet{Strauss2002}, but used less stringent thresholds in terms of Petrosian magnitude cut, star-galaxy separation, fiber magnitude cut, and rejection of bright objects. We therefore re-implement the stricter thresholds of \citet{Strauss2002} and we select, as before, sources with good redshift measurements (\textsc{zwarning} = 0, $z > 0$). This leaves us with a sample of $685\,813$ galaxies which we match in position to the Legacy MGS with a tolerance of $0.5 \arcsec$. We check that the redshift of the matched objects is consistent and we identify $252\,965$ galaxies with measured K-band magnitude. 

\subsection{The numerical simulation}
\label{numsim}
Constrained simulations are numerical simulations whose initial conditions are constrained from a sample of observed galaxies. These initial conditions, evolved with a N-body code, result in a simulation that reproduces the existing structures of the Local Universe. The simulation for this work has been developed following the technique detailed in \citet{Sorce2016} and \citet{Sorce2018}. The initial conditions are created from catalogues of distances to galaxies and groups \citep{Tully2013, SorceTempel2017} which are converted to peculiar velocities following \citet{SorceTempel2018, Sorce2016vel} and bias-minimized \citep{Sorce2015}. We stress that only the information on the peculiar velocities is used to constrain the initial conditions, without any use of the observed density field information.

The dark matter-only simulation used in this work was built on initial conditions containing $2048^3$ particles in a box of side $500 \mathrm{Mpc}/h$. The dark matter particle mass resolution offered by the simulation is of $10^{9} M_{\sun}/h$, which allows us to resolve dark matter halos of $10^{11} M_{\sun}/h$ with 100 particles at $z = 0$. The simulation was run from $z = 120$ to $z = 0$ with the adaptive mesh refinement code \textsc{Ramses} \citep{Teyssier2002}. The best-achieved spatial resolution is of $\sim 1.89 \mathrm{kpc}/h$ thanks to subsequent refinements of the mesh (a level is refined if the total density in a cell is larger than that of a cell containing 8 dark matter particles).

We identify dark matter halos and subhalos with the \textsc{Halomaker} software \citep{Aubert2004, Tweed2009}, modified to operate with $2048^{3}$ (i.e. more than $2^{31}$) particles. Dark matter halos are identified in real space with the local maxima of the dark matter particle density field. Their boundary is defined as the point where the dark matter mass over-density is lower than 80 times the background density. The result is a sample of $3\,666\,018$ dark matter halos in the simulation box.

\subsection{Identifying the galaxy population in the numerical simulation}
\label{idgalinsim}
We identify a suitable subset of halos to mimic our galaxy population in two ways: with a mass cut and with an apparent K-band magnitude cut. In the rest of this paper we refer to these halos as "galaxies". We do not make a distinction between whether a halo is a main halo or the sub-halo of a main one. We use these halos to run \disperse~to detect the cosmic web. Our mass-selected galaxy sample is composed of all $749\,960$ halos with mass in the range $10^{12} \leq M_{\mathrm{vir}} (M_{\sun}) \leq 10^{13}$. To create our magnitude-selected galaxy sample we first computed the K-band absolute magnitude for every halo in the simulation by setting it proportional to the circular velocity of each halo (i.e. by inverting the Tully-Fisher relation, as obtained with fits to observed or simulated samples, see e.g. Equation 7 of \citealt{Mathis2002} and Equation 3 of \citealt{VdBosch2000}):
\begin{equation}\label{kmageq}
M_{K} = -2.34-8.16 \log(2V_{\mathrm{circ}})
\end{equation}
The numerical values in this relation have been provided by K. Dolag (private communication). The absolute K-band magnitude is then converted to apparent magnitude by means of the distance modulus: $m_{K} = M_{K} + 25+5\log(d)$. Based on our analysis of the K-band magnitude of SDSS galaxies we set a threshold at $m_{K} = 14$ and select all halos brighter than this limit ($263\,272$) as galaxies.

\begin{figure}
\centering
\resizebox{\hsize}{!}{\includegraphics{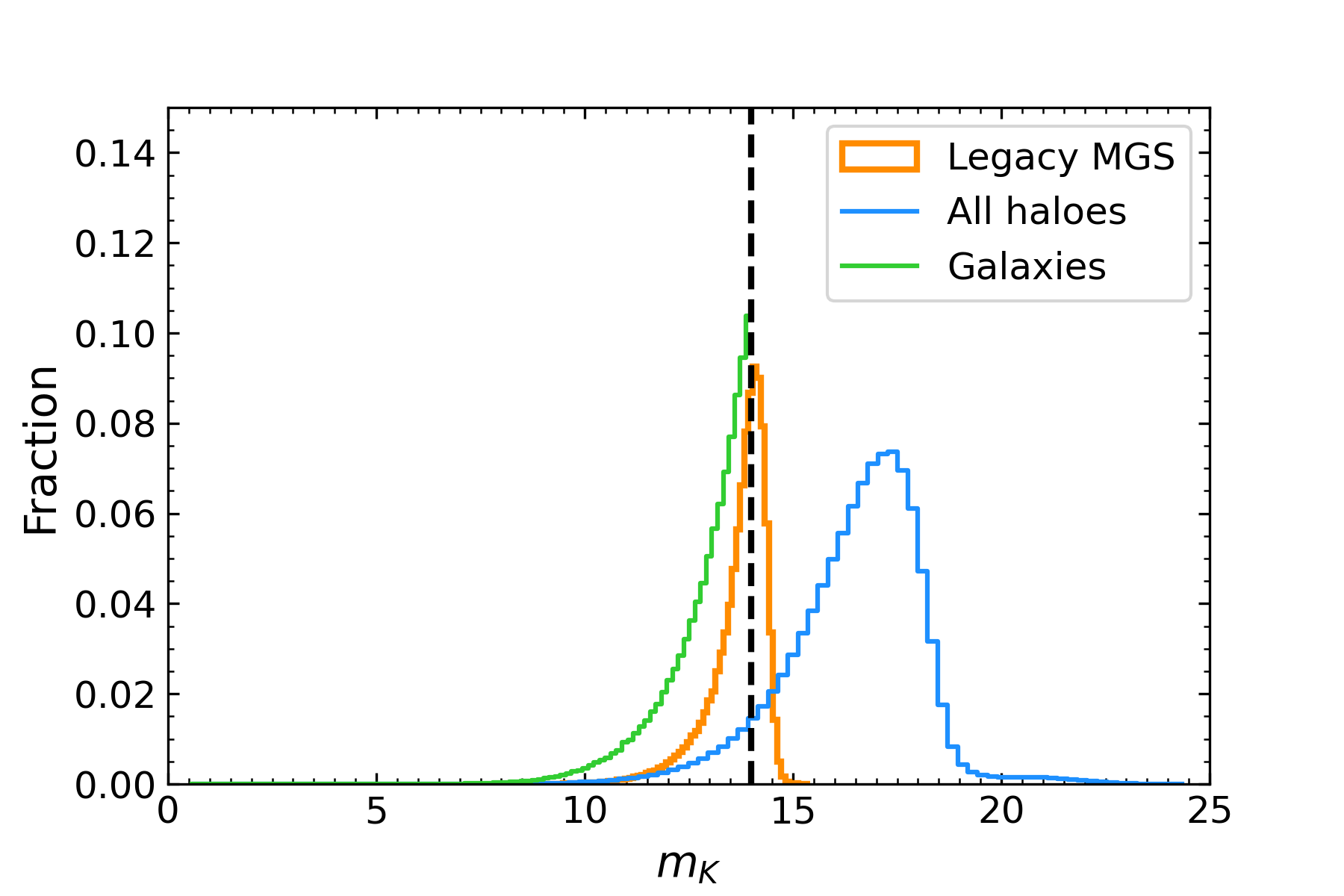}}
\caption{Apparent K-band magnitude distributions for the halos in our simulation and for SDSS galaxies. The orange line refers to the Legacy MGS galaxies with apparent K-band magnitude from the NYU-VAGC sample. The light blue line refers to the $m_{K}$ distribution for all halos in our simulation (with $m_{K}$ computed with Eq. \eqref{kmageq}). The green line shows the distribution for our magnitude-selected simulated galaxy sample. The vertical black dashed line shows our magnitude threshold of $m_{K} = 14$.}
\label{kdist}
\end{figure}

Figure \ref{kdist} shows the distribution of the apparent K-band magnitude for all halos in the simulation (as computed with Eq. \eqref{kmageq}), those selected as galaxies and for galaxies in the Legacy MGS sample (by matching with the NYU-VAGC, as explained in Section \ref{obsdata}). The distribution of Legacy MGS galaxies sharply drops around the value we selected as magnitude limit for our simulated galaxy sample, therefore providing supporting evidence for our choice. The distribution for all halos, however, peaks at much fainter magnitudes (due to the simulation being complete to the resolution limit for the halos), resulting in us selecting a small fraction of the available halos.

We note that our magnitude selected sample of simulated galaxies includes halos with large values of $M_{\mathrm{vir}}$ (up to $M_{\mathrm{vir}} \sim 10^{14} M_{\sun}$). This is visible, for example, in the mass distribution of magnitude selected galaxies shown in Appendix \ref{appendix_masses}, Figure \ref{mdist}. As our galaxy samples include both main and sub-halos, we identify this population of galaxies residing in high-mass halos to be galaxies at the center of groups and clusters (e.g. BCGs). Indeed, deriving the mass distribution of only those magnitude selected simulated galaxies that are not main halos, it shifts to lower masses, more consistent with observations (not shown here). Given that BCGs are not eliminated from the Legacy MGS either, that the K-band apparent magnitude distributions of real and simulated galaxies are in agreement, and that \disperse~is run only considering galaxy positions (without information about their mass or K-band magnitude required), we do not consider this to be a problem in our detection of the cosmic web.

Conversely, we note that the presence of an upper mass limit of $10^{13} M_{\sun}$ in our mass-selected simulated galaxy sample largely excludes these BCGs in massive halos from the sample. To understand whether this may be a problem, we have performed our analysis also including these BCGs in massive halos in our mass-selected galaxy sample (in the number of $85\,953$ halos of mass $M_{\mathrm{vir}} > 10^{13} M_{\sun}$ added to those already present in the mass-selected galaxy sample), i.e. by running \disperse~on a sample of simulated galaxies selected to have $M_{\mathrm{vir}} \geq 10^{12} M_{\sun}$. We do not find substantial differences with the conclusions reported in this paper with our current definition of a mass-selected galaxy sample ($10^{12} \leq M_{\mathrm{vir}} (M_{\sun}) \leq 10^{13}$) and we do not show the results of this additional test in the paper. In the following, we will use a mass-selected galaxy sample with both a lower and upper mass limit to highlight how the cosmic web in the vicinity of Coma is relatively insensitive of our definition for a galaxy population (as long as it is reasonably similar to the Legacy MGS). We postpone to future papers the task of identifying an even more realistic simulated galaxy sample in our simulation through semi-analytical models.

\subsection{Identifying the Coma cluster in the numerical simulation}
\label{identifyingcoma}
We identify the halo in our simulation corresponding to the Coma cluster following \citet{Sorce2018}. First we select all halos more massive than $10^{14.5} M_{\sun}$ and whose distance on the Line-of-Sight (LoS) is within $\pm 30 \%$ of the distance to the Coma cluster, i.e. $|d_{h}-d_{\mathrm{Coma}}|/d_{\mathrm{Coma}} \leq 0.3$ where $d_{h}$ is the distance to the considered halo and $d_{\mathrm{Coma}}$ is the distance to the Coma cluster. We derive $d_{\mathrm{Coma}} = 68.6 \mathrm{Mpc}/h$ by computing the comoving distance corresponding to $z = 0.023$ (which is the redshift of the Coma cluster adopted for the analysis of \citealt{Malavasi2020Coma}, see Table 1). Of the three candidates thus selected, we identify Coma with the most massive, which is also the one whose coordinates are the closest to the coordinates of the real Coma. In the rest of the paper we will refer to this halo as "simulated Coma" and to the real cluster as "real Coma". We note that there is still a residual difference in the supergalactic coordinates of the real and simulated Coma of $\sim 7 \div 14\: \mathrm{Mpc}/h$. In the following, we work under the assumption that this residual difference in the coordinates of the simulated and real Coma is due to the simulation and not to the astrometric uncertainty in the coordinates of the real Coma, which we assume to be correct. We therefore apply a corresponding shift to the reference frame of the simulated box (i.e. to the coordinates of all halos including the one corresponding to Coma and of simulated galaxies) so that the position of the simulated and real Coma are identical. As this shift does not distort the box nor change its dimensions, it does not affect our conclusions. The physical parameters of the simulated Coma are therefore $M_{\mathrm{vir}} = 1.259 \cdot 10^{15} M_{\sun}/h$, $M_{200} = 9.441 \cdot 10^{14} M_{\sun}/h$, and $r_{\mathrm{vir}} = 1.75 \mathrm{Mpc}/h$. We report a value for the mass which is higher for the simulated Coma than for the real one used in \citet[][real Coma: $M_{200} = 3.57 \cdot 10^{14} M_{\sun}$, from \citealt{Gavazzi2009}, $M_{\mathrm{vir}} = 9.8 \cdot 10^{14} M_{\sun}/h$, from \citealt{LokasMamon2003}]{Malavasi2020Coma}. This difference could be due to the performance of the halo and sub-halo detection software, which may include in the mass determination for the halo corresponding to the simulated Coma unresolved substructure or particles which are actually unbound (see e.g. \citealt{Knepe2011} for a comparison of halo detection algorithms), or to uncertainties in the Coma mass determination from observations. The mass of our simulated Coma is in better agreement with (albeit slightly lower than) the mass for the real Coma cluster reported by \citet[][$M_{\mathrm{vir}} = 1.7 \cdot 10^{15} M_{\sun}/h$]{Tully2015}. In our case, the difference between the mass of the real and simulated Coma is relevant when addressing the connectivity-mass relation in Section \ref{simcomaconn}. Indeed this difference in the mass prevents us from decisively concluding whether the connectivity measurement for the simulated Coma is in agreement with the observed value. However, including the uncertainty on the observed mass measurement, we can conclude that the connectivity of simulated Coma is broadly in line with the expected trends from observations and numerical simulations in the literature (see Section \ref{simcomaconn}).

\section{Inclusion of the redshift space distortions}
\label{foginclusion}
In order to perform a more realistic comparison between the cosmic web we detect in simulations and the one we detect in observations, we have introduced redshift space distortions (Finger of God effect, FoG) in the position of our simulated galaxies. The FoG effect is a distortion of galaxy positions along the LoS in the vicinity of massive structures due to the fact that galaxy peculiar velocities introduce an error in the measurement of their redshift \citep{Kaiser1987}. Due to this effect, clusters appear as elongated structures (and not spherical ones) in the LoS direction, and they can be easily mistaken for a filament by algorithms such as \disperse. As the filaments detected around Coma in observations are recovered with the FoG effect being present in the Legacy MGS \citep[see the discussion of this point in][]{Malavasi2020Catalogue, Malavasi2020Coma} while this effect is absent in our constrained simulation, we introduced it in order to make the comparison more realistic.

We introduced the FoG effect in two ways, both of which follow a general procedure: first of all we computed peculiar velocities for all the galaxies in our simulation, then we converted peculiar velocities to redshift, and finally we added the peculiar redshift to the cosmological one for each galaxy. The peculiar velocity ($\boldsymbol{v}$) is provided in cartesian supergalactic coordinates for each galaxy as a direct output of our simulation (see also Section \ref{accretion}). We convert it into supergalactic spherical coordinates, which provides us with a proper motion for each galaxy on the plane of the sky (which we ignore) and a radial velocity (to which we refer as $v_{\mathrm{pec}}$ in the following). This velocity is along the LoS (negative or positive depending whether a galaxy is approaching or receding). While this radial velocity is directly provided by the simulation, we also obtained $v_{\mathrm{pec}}$ in a different way for completeness and to test a different method: for each galaxy which is a main halo (i.e. only for halos which are not substructures of other main halos) we derive the velocity dispersion from the virial mass, using the formula $\sigma_{v} = \sqrt{GM_{\mathrm{vir}}/5r_{\mathrm{vir}}}$ (see also \citealt{Malavasi2020Catalogue}). For each sub-halo (i.e. only for halos which are substructures of other main halos), we then randomly extracted a value of $v_{\mathrm{pec}}$ from a Gaussian distribution centered on zero and with a dispersion of $\sigma_{v}$. As this approach of determining $v_{\mathrm{pec}}$ provided results consistent with using the peculiar velocities directly provided by the simulation, we will not discuss it further and we will rather focus on the first method.

Once a value of $v_{\mathrm{pec}}$ was determined, we converted it to redshift using the formula $z_{\mathrm{pec}} = v_{\mathrm{pec}}/c$, with $c$ being the speed of light. We then added this peculiar redshift to the true redshift of each galaxy (as derived from its distance, obtained from its supergalactic coordinates) as described in \citet{Tully2013} and \citet{Sorce2023velwaves}: $z_{\mathrm{obs}} = (1+z_{\mathrm{true}})(1+z_{\mathrm{pec}})-1$. This observed redshift ($z_{\mathrm{obs}}$) was then used to compute a new distance for each galaxy, which was then used to compute new supergalactic cartesian coordinates (assuming the angular position of each galaxy on the plane of the sky to remain unchanged). These new positions were then used to run \disperse in the same way we did for the galaxy sample where the FoG effect was not introduced (see Section \ref{dispdesc}). We visually checked that the prominence of the FoG for the simulated Coma cluster constructed in this way is consistent with the FoG for the observed Coma present in both the Legacy MGS and the sample of galaxies of \citet{Tully2013}.

\section{Detecting the cosmic web}
\label{dispdesc}
We detect the cosmic web in the simulation by applying the same cosmic web detection algorithm used in \citet{Malavasi2020Catalogue, Malavasi2020Coma} to our population of simulated galaxies.

The Discrete Persistent Structure Extractor (\disperse, \citealt{Sousbie2011a, Sousbie2011b}) extracts the filaments of the cosmic web by means of their topology. It works on discrete sets of points: galaxy distributions from surveys, and halo distributions in simulations. For the application of \disperse~to the Legacy MGS we refer the reader to \citet{Malavasi2020Catalogue}. In this work, we apply \disperse~to our galaxy populations selected through the mass and K-band magnitude cuts. The starting point is a measurement of the galaxy density field, obtained through the Delaunay Tessellation Field Estimator algorithm \citep{SchaapVdW2000, VdWSchaap2009}. While not necessary for the filament extraction, the galaxy density field can be smoothed by recursively averaging the density measured at the position of each galaxy with the density of all galaxies connected to it by edges of the tetrahedrons of the Delaunay tessellation. For our analysis we will work with either an un-smoothed density field (referred to as "SD0") or a density field smoothed only once ("SD1"). \disperse~then computes the discrete gradient of the density field and identifies points (called critical points) where the gradient is null (this is done by applying the discrete Morse theory). These critical points can be maxima, minima, and two type of saddles (1-saddles, local density minima bound to walls, and 2-saddles, local density minima bound to filaments). Filaments consistently connect maxima and 2-saddles together through lines of constant gradient (ridges of the density field). The \disperse~algorithm implements a selection based on the persistence of the filaments (tied to the density contrast of the critical points at their extrema and measured based on the persistent homology theory). The distribution of persistence values for the pairs of critical points detected in the data is compared to the same distribution for pairs of critical points detected in a Gaussian random field (which models the set of filaments and critical points which would be obtained if only noise and no real data were present). All filaments connected to critical points closer (in terms of their persistence value) to the noise distribution than a certain number of standard deviations are removed via topological simplification. In this work we consider persistence thresholds for the elimination of spurious filaments of $2\sigma$ and $3\sigma$. The skeleton (i.e. the set of filaments and critical points) thus constructed is then post-processed: artificial critical points (called bifurcations) are inserted where two or more filaments intersect without a maximum being present. The filaments are composed of small segments (the size of the edges of the tetrahedrons of the Delaunay tessellation). The filament shape is smoothed by recursively averaging the positions of the extremes of each of these segments with the positions of the extremes of the segments directly attached to it (without modifying the positions of the critical points at the extremes of the filaments). In this work, this smoothing is done only once.

\section{Comparing the cosmic web in simulated and observed data} 
\label{cwcomparison}
The approach to detect the cosmic web used in this work is the same as the one used in \citet{Malavasi2020Coma}. As a reference, from observations we select the skeleton extracted in the Legacy MGS with the combination of parameters SD1 (1 smoothing cycle) and a persistence threshold of $3\sigma$ (see the top panel of Figure 2 of \citealt{Malavasi2020Coma}). The first step in our analysis is represented by finding among the parameter combinations explored in this work (\disperse~parameters and galaxy mass vs. magnitude selection) those which yield the best match between the skeleton detected in the constrained simulation and the one detected in observations. As a perfect match between the simulation and observations is very unlikely due to the residual cosmic variance of the former\footnote{i.e. we cannot directly constrain the initial conditions below the non-linear/linear threshold with the current technique.}, we aim at excluding some of the parameter combinations by means of this comparison. In the rest of this section we proceed by first comparing visually the simulated and observed skeleton, then by comparing some global properties of the filaments and critical points and finally by comparing some simulated and observed properties of the Coma cluster and its filaments. We begin by comparing the observed skeleton with the one derived in the constrained simulation without the inclusion of the FoG effect. We provide a comparison with the inclusion of the FoG effect in Section \ref{subsfog}.

\subsection{Visual comparison of the filament configuration}
We start by identifying the critical points associated with Coma, in a similar way to what we did in \citet{Malavasi2020Coma}. For each combination of \disperse~parameters, we identify all the critical points within 1 $r_{\mathrm{vir}}$ of the simulated Coma. Table \ref{cp_inside_coma} reports our findings, broken down by critical point type, as well as our results from the parameter combination $\mathrm{SD1} - 3\sigma$ from Table 2 of \citet{Malavasi2020Coma}, for reference.

\begin{table*}
\caption{Critical points associated with Coma. The first four columns refer to the critical points detected within the virial radius of the simulated Coma with the various combinations of \disperse~parameters. The last column provides the critical points detected within the virial radius of the real Coma in \citet{Malavasi2020Coma}.}           
\label{cp_inside_coma}   
\centering                                   
\begin{tabular}{c c c c c | c}         
\hline\hline                       
\multicolumn{6}{c}{Mass selected galaxies} \\                         
\hline
CP type & $\mathrm{SD0} - 2\sigma$ & $\mathrm{SD0} - 3\sigma$ & $\mathrm{SD1} - 2\sigma$ & $\mathrm{SD1} - 3\sigma$ & \citet{Malavasi2020Coma} \\   
\hline
Minima 	   & 0 & 0 & 0 & 0 & 0 \\  
1-saddles    & 0 & 0 & 0 & 0 & 0 \\  
2-saddles    & 0 & 0 & 0 & 0 & 0 \\  
Maxima  	   & 1 & 1 & 1 & 1 & 1 \\ 
Bifurcations & 0 & 0 & 1 & 0 & 1 \\
\hline                                        
\multicolumn{6}{c}{Magnitude selected galaxies} \\                           
\hline
CP type & $\mathrm{SD0} - 2\sigma$ & $\mathrm{SD0} - 3\sigma$ & $\mathrm{SD1} - 2\sigma$ & $\mathrm{SD1} - 3\sigma$ & \citet{Malavasi2020Coma} \\   
\hline
Minima 	   & 0 & 0 & 0 & 0 & 0 \\  
1-saddles    & 1 & 1 & 0 & 0 & 0 \\  
2-saddles    & 2 & 1 & 0 & 0 & 0 \\  
Maxima  	   & 1 & 1 & 1 & 1 & 1 \\ 
Bifurcations & 1 & 1 & 0 & 0 & 1 \\
\hline
\end{tabular}
\end{table*}

Table \ref{cp_inside_coma} shows how the simulated Coma cluster is consistently identified with a maximum regardless of the combination of \disperse~parameters or galaxy selection method. In the case of the mass selection for the simulated galaxy population, an additional bifurcation point is found inside $r_{\mathrm{vir}}$, similar to what happens in the case of \citet{Malavasi2020Coma} in some of the \disperse~parameter combinations. In the case of magnitude-selected galaxies, the amount of critical points detected within the $r_{\mathrm{vir}}$ of Coma is larger. When no smoothing and a low persistence threshold are chosen ($\mathrm{SD0} - 2\sigma$) both 1-saddles and 2-saddles (local minima bound to filaments and sheets) are also found within the $r_{\mathrm{vir}}$ of Coma.

\begin{figure*}
\sidecaption
\includegraphics[trim = 0.75cm 4cm 1cm 4.2cm, clip = true, width = 12cm]{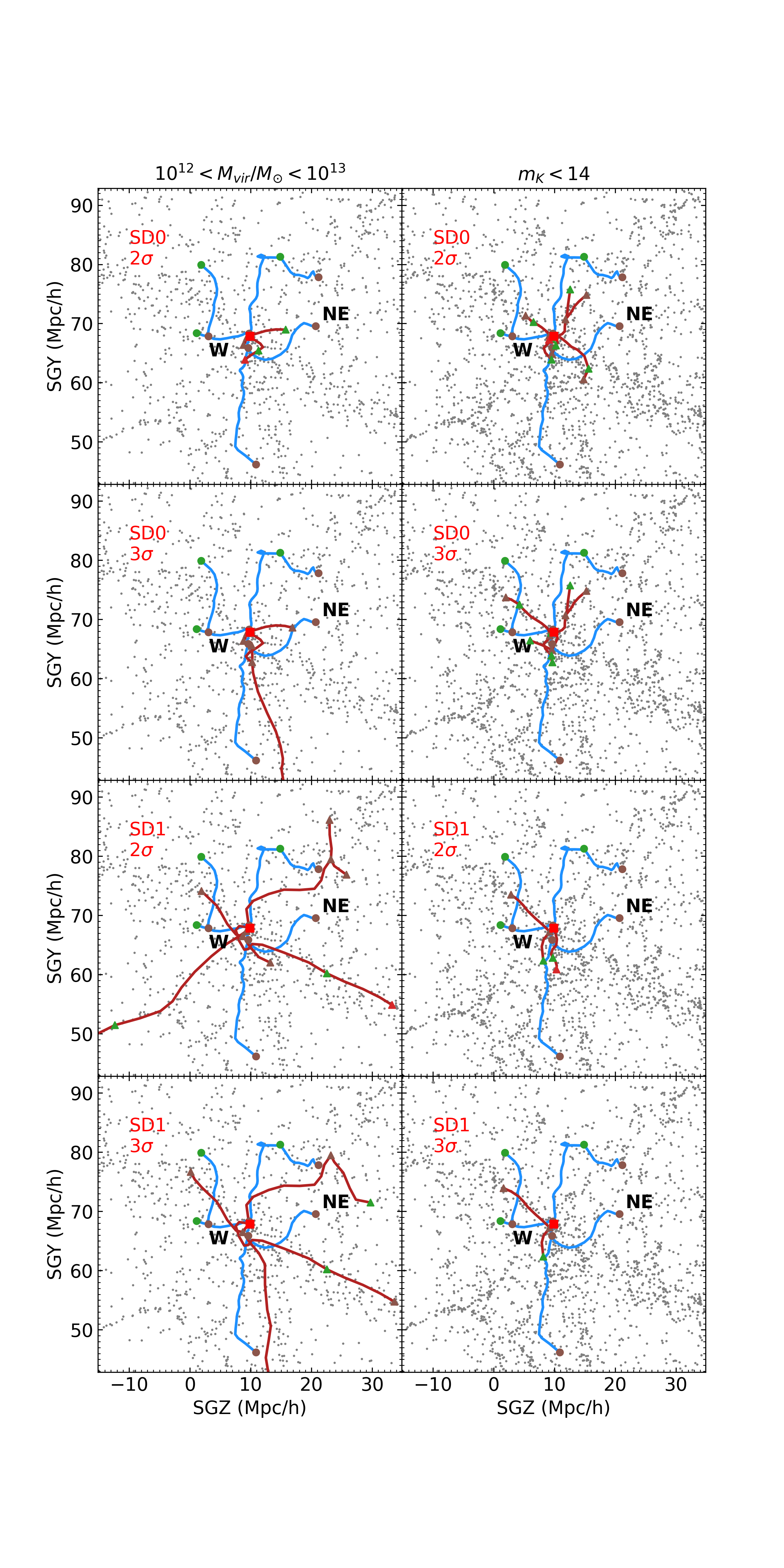}
\caption{Filaments in the constrained simulation compared to filaments in observations. The left column shows the case of a mass selected galaxy sample in the constrained simulation, the right column shows the case of a magnitude selected galaxy sample (see text). In each panel, simulated galaxies are shown as grey points in a slice of thickness 50 Mpc/h centered on Coma. Blue lines are filaments from \citet{Malavasi2020Coma} in supergalactic coordinates, the red square marks the position of the Coma cluster. Dark red lines are filaments obtained in the simulation with a variety of \disperse~parameter combinations, namely: $\mathrm{SD0} - 2\sigma$ (top row), $\mathrm{SD0} - 3\sigma$ (second row), $\mathrm{SD1} - 2\sigma$ (third row), $\mathrm{SD1} - 3\sigma$ (bottom row). The observed NE and W filaments are marked for reference. Circle and triangles mark the positions of critical points (in observations and simulations, respectively) and are color-coded according to their type (red: maxima, green: type 2 saddles, brown: bifurcations). This figure shows the SGZ-SGY projection.}
\label{visual_skel_comparison_proj1}
\end{figure*}

Starting from this selection of critical points we recursively find all the filaments attached to them and those attached to the critical points at their ends, iterating the search for filaments up to a radius of 75 comoving Mpc (50.8 Mpc/$h$), as it was done in \citet{Malavasi2020Coma}. In agreement with that work, we call the filaments connected to the critical points within the virial radius of Coma "first generation" and those connected to the critical points at the other end of first generation filaments (and not within the virial radius of Coma) "second generation". "Third generation" filaments are then connected to the critical points at the other end of second generation filaments (and which are not connected to first generation filaments), "fourth generation" filaments are connected to third generation, etc. We show first and second generation real and simulated filaments in Figure \ref{visual_skel_comparison_proj1}. This figure shows a supergalactic coordinate projection (SGZ-SGY) in which first and second generation filaments connected to the real and simulated Coma are compared. In this figure, the redshift direction is aligned with the SGY axis, while the SGZ axis is located in the plane of the sky. The NE and W filaments from the observations (in blue) are highlighted in this plot, clearly visible intersecting at the cluster position in an almost straight axis (in the proximity of the cluster). Perpendicular to this axis is an elongated filament which is located roughly along the Finger of God (FoG) due to the Coma cluster. While such a filament is most probably unphysical and the result of the FoG distortion, in \citet{Malavasi2020Coma} we performed tests to check whether the fact of not having compressed the FoGs had any bearing on our results related to the connection of the filaments to the cluster, while in \citet{Malavasi2020Catalogue} we demonstrated that the filament reconstruction in the SDSS, without compressing the FoG, did not statistically affect the analysis we intended to perform. In this work we aim to identify the match between observations and simulations on the basis of the positions of the NE and W filaments alone, whose detection is more secure as outlined in \citet{Malavasi2020Coma}. Given the observed filaments' orientation (i.e. the fact that the NE and W filaments are along an almost-straight axis along the SGZ direction), we believe this is the best projection to identify possible matches between observed and simulated filaments. We provide an extra projection in Appendix \ref{extraproj} along with a short discussion.

Several of the \disperse~parameter combinations offer filament configurations around Coma which are similar to what observed in \citet{Malavasi2020Coma}. In particular, $\mathrm{SD0} - 2\sigma$ in the case of magnitude selected galaxies and $\mathrm{SD1} - 2\sigma$ and $\mathrm{SD1} - 3\sigma$ in the case of mass selected galaxies all show prominent NE-W filaments aligned in an axis at around the same location as the observed ones. In the case of magnitude selected galaxies, the W filament is present in all \disperse~parameter configurations, while the NE filament is present also in the case of $\mathrm{SD1} - 2\sigma$, although shorter in length.

Based on these considerations, it seems that in the case of magnitude selected galaxies, the most important \disperse~ parameter is the persistence threshold (which has to be of the order of $2\sigma$), while in the case of the mass selected galaxies smoothing seems the most important parameter.

\subsection{Statistical comparison of the skeleton properties}
We then check how the selected skeletons compare in terms of their general properties. We stress that this comparison is less constraining in terms of determining which simulated skeleton is similar to the observed one as several statistical quantities of the skeleton (e.g. the density of the critical points and the length of the filaments) will strongly depend on the density of tracers (higher in the simulations and dependent on redshift for the observations) and on the region in which the skeleton is extracted. However, such comparisons (in particular for the connectivity) may still be interesting to understand better the differences between simulated and observed skeleton and to have a general view of the properties of filaments explored.

\begin{figure}
\centering
\resizebox{\hsize}{!}{\includegraphics{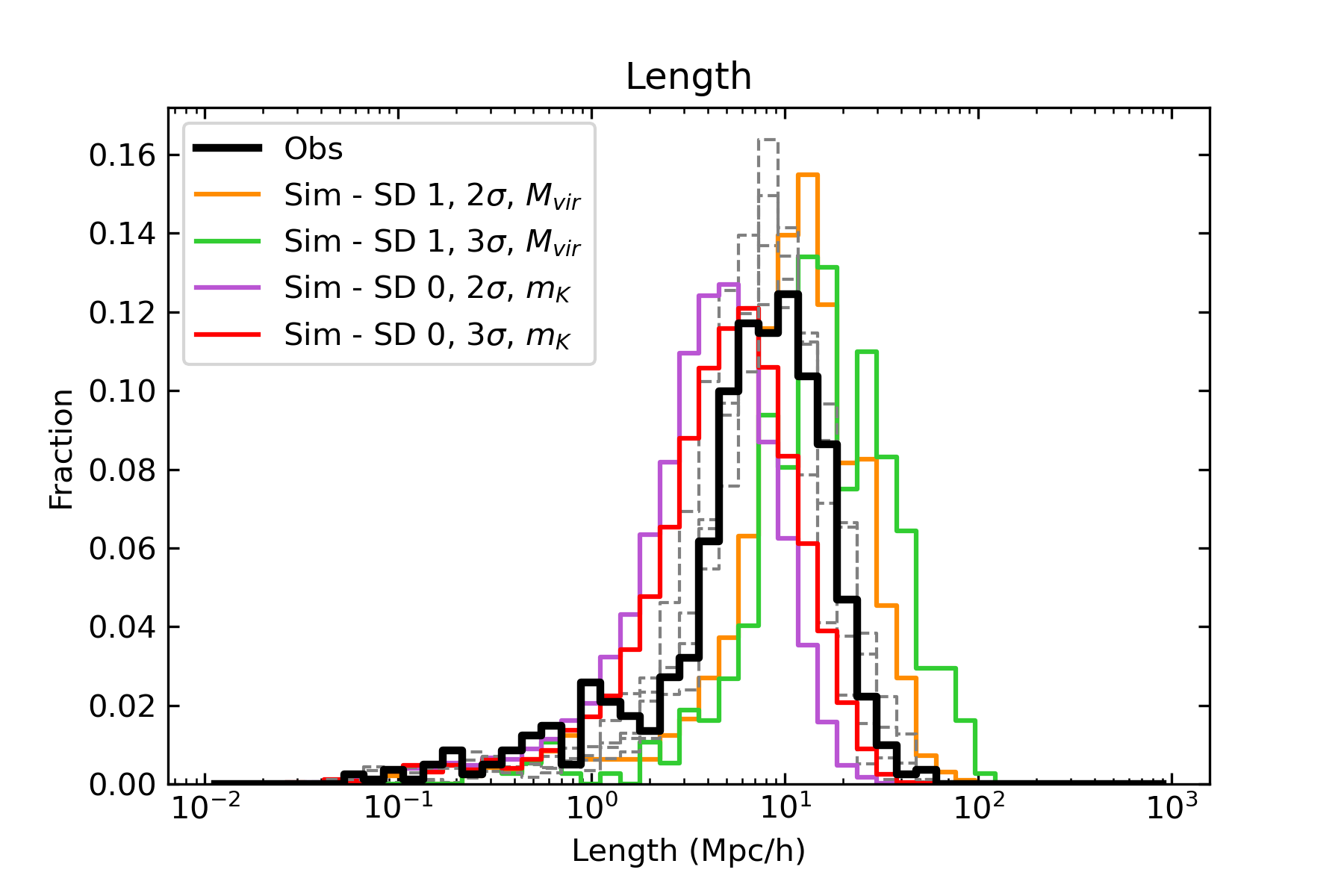}}
\caption{Filament length distribution. The black bold distribution refers to observed filaments from \citet{Malavasi2020Coma} with $\mathrm{SD1} - 3\sigma$. The colored lines refer to simulated filaments with parameter combinations $\mathrm{SD0} - 2\sigma$ and $\mathrm{SD0} - 3\sigma$ (for magnitude selected galaxies, in purple and red, respectively), and $\mathrm{SD1} - 2\sigma$ and $\mathrm{SD1} - 3\sigma$ (for mass selected galaxies, in orange and green, respectively). Dashed grey lines represent the rest of the parameter combinations and provide an idea of the range of lengths explored. Only filaments within a slice of thickness $\pm 75 \mathrm{Mpc}$ (50.8 Mpc/$h$) from the location of the (simulated or observed) Coma cluster were used to derive the distributions.}
\label{lendist}
\end{figure}

In Figure \ref{lendist} we begin by showing the distributions of filament lengths for the observed and simulated skeletons. To derive this and the following distributions, both for observations and simulations, we have focused only on filaments and critical points within a slice of thickness $\pm 75 \mathrm{Mpc}$ (50.8 Mpc/$h$) from the location of the (simulated or observed) Coma cluster. We do so in order to limit in redshift the filaments from the observations for which we derive properties (and ensure that they are extracted in a region where the density of tracers is more or less constant) and to limit in density the region we explore (the environment of a rich cluster will be different from the average density field and the filaments there may not be representative of the average filament in the field we explore).

The observed and simulated skeletons have largely similar length distributions, showing that although the tracer density may differ, the filaments have comparable lengths. In this and the following Figures, we highlight with colored lines the distributions corresponding to the four simulated skeletons selected as those that best match the observations via visual comparison in the previous step (i.e. the combinations $\mathrm{SD0} - 2\sigma$ and $\mathrm{SD0} - 3\sigma$ for magnitude selected galaxies and $\mathrm{SD1} - 2\sigma$ and $\mathrm{SD1} - 3\sigma$ for mass selected galaxies). For completeness, we report in grey the distributions corresponding to the simulated skeletons obtained with other parameter combinations not selected via visual comparison in Figure \ref{visual_skel_comparison_proj1}. We do so to show the full ranges of skeleton properties explored by our simulated skeletons while highlighting the \disperse~parameter combinations for the simulated skeletons that we use in our analysis. Figure \ref{lendist} shows how these distributions are actually those with the largest systematic difference in terms of filament length distributions from the observed skeleton. Simulated skeletons with lower values of smoothing (SD0) are characterized by systematically shorter filaments than observations, while skeletons with larger values of smoothing (SD1) have filaments which are on average longer\footnote{This behavior is expected, see e.g. Figure 12 of \citet{Malavasi2020Catalogue}.}. Our selection therefore allows to explore two extrema in the relation between observed and simulated filaments (at least with respect to their length distribution) and to obtain a better idea of the general behavior.

\begin{figure}
\centering
\resizebox{\hsize}{!}{\includegraphics{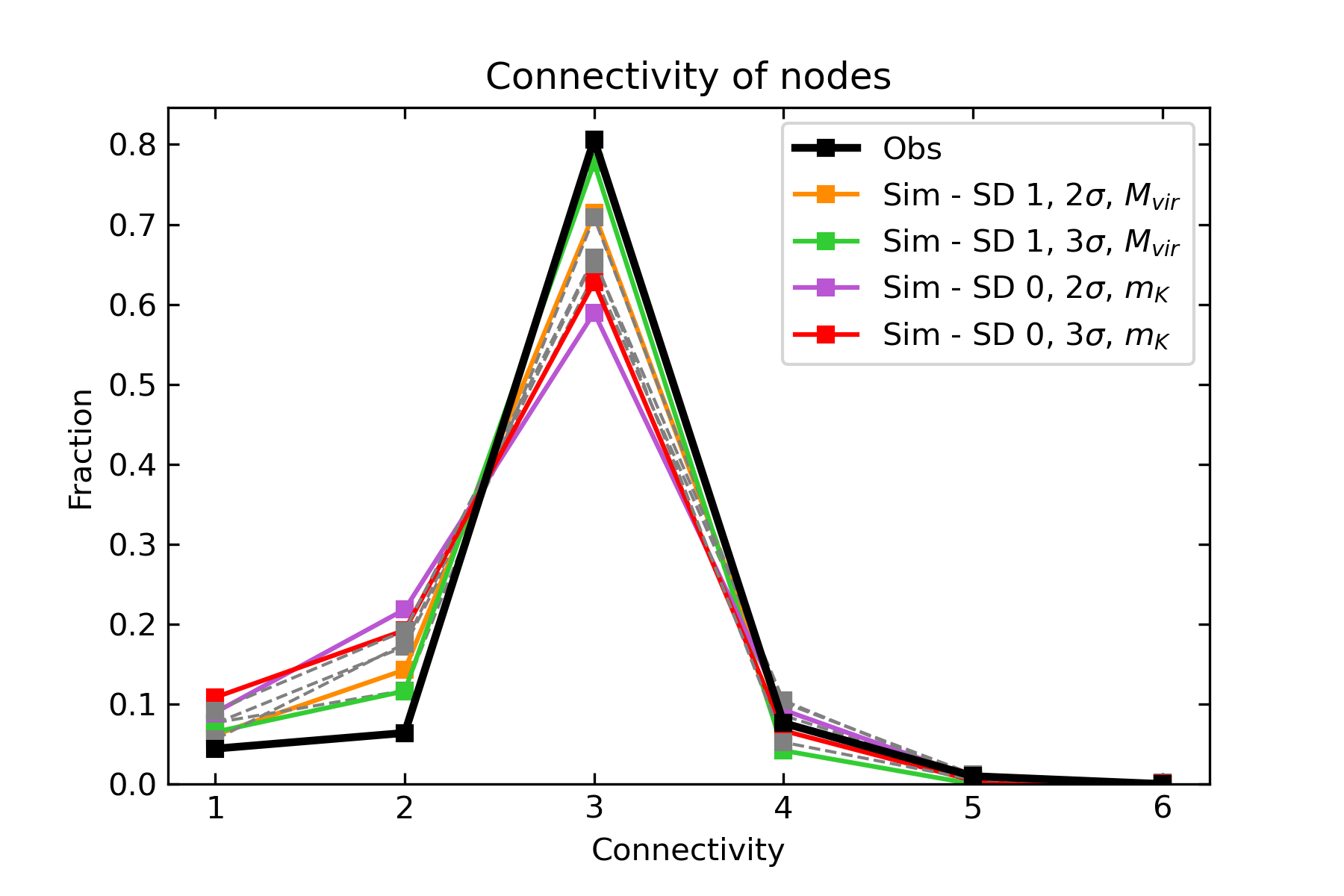}}
\caption{Node connectivity distribution. The black bold distribution refers to observed nodes from \citet{Malavasi2020Coma} with $\mathrm{SD1} - 3\sigma$. The colored lines refer to simulated nodes with parameter combinations $\mathrm{SD0} - 2\sigma$ and $\mathrm{SD0} - 3\sigma$ (for magnitude selected galaxies, in purple and red, respectively), and $\mathrm{SD1} - 2\sigma$ and $\mathrm{SD1} - 3\sigma$ (for mass selected galaxies, in orange and green, respectively). Dashed grey lines represent the rest of the parameter combinations and provide an idea of the range of connectivities explored. Only nodes within a slice of thickness $\pm 75 \mathrm{Mpc}$ (50.8 Mpc/$h$) from the location of the (simulated or observed) Coma cluster were used to derive the distributions.}
\label{conndist}
\end{figure}

We then focus on the connectivity of maxima and bifurcations. In this case, we define the critical point connectivity as the number of filaments connected to a given maximum or bifurcation. We explore the connectivity distribution for filaments in the same distance range of $\pm 75 \mathrm{Mpc}$ (50.8 Mpc/$h$) around Coma as for the length distribution. The connectivity distribution for maxima and bifurcations combined is shown in Figure \ref{conndist}. In the following of this paper we refer to maxima and bifurcations together as nodes. We stress that this term is not used in a topological sense (unlike maxima and bifurcations), but in an astrophysical one to refer to dense structures of the cosmic web.

This Figure shows that the majority of points in the vicinity of Coma have a connectivity of three and this is consistent between observations and simulations. For reference, for the Coma cluster, \citet{Malavasi2020Coma} reported a connectivity of 2.5. All the \disperse~parameter combinations we explored are in agreement with the observed connectivity distribution, having a peak at a value of $\kappa = 3$. This figure shows the same situation as the length distribution: the parameter combinations we selected based on visual identification (highlighted in color in the Figure) are at the outskirts of the range covered by the various connectivity distributions from the simulations. Parameter combinations with a larger smoothing (SD1) have a larger fraction of nodes with a connectivity of three (the combination $\mathrm{SD1} - 3\sigma$ for mass selected galaxies is the closest to the observed distribution), while combinations with a lower smoothing have a lower fraction of nodes with a connectivity of three and a larger fraction of nodes with a connectivity of two.

\begin{figure}
\centering
\resizebox{\hsize}{!}{\includegraphics{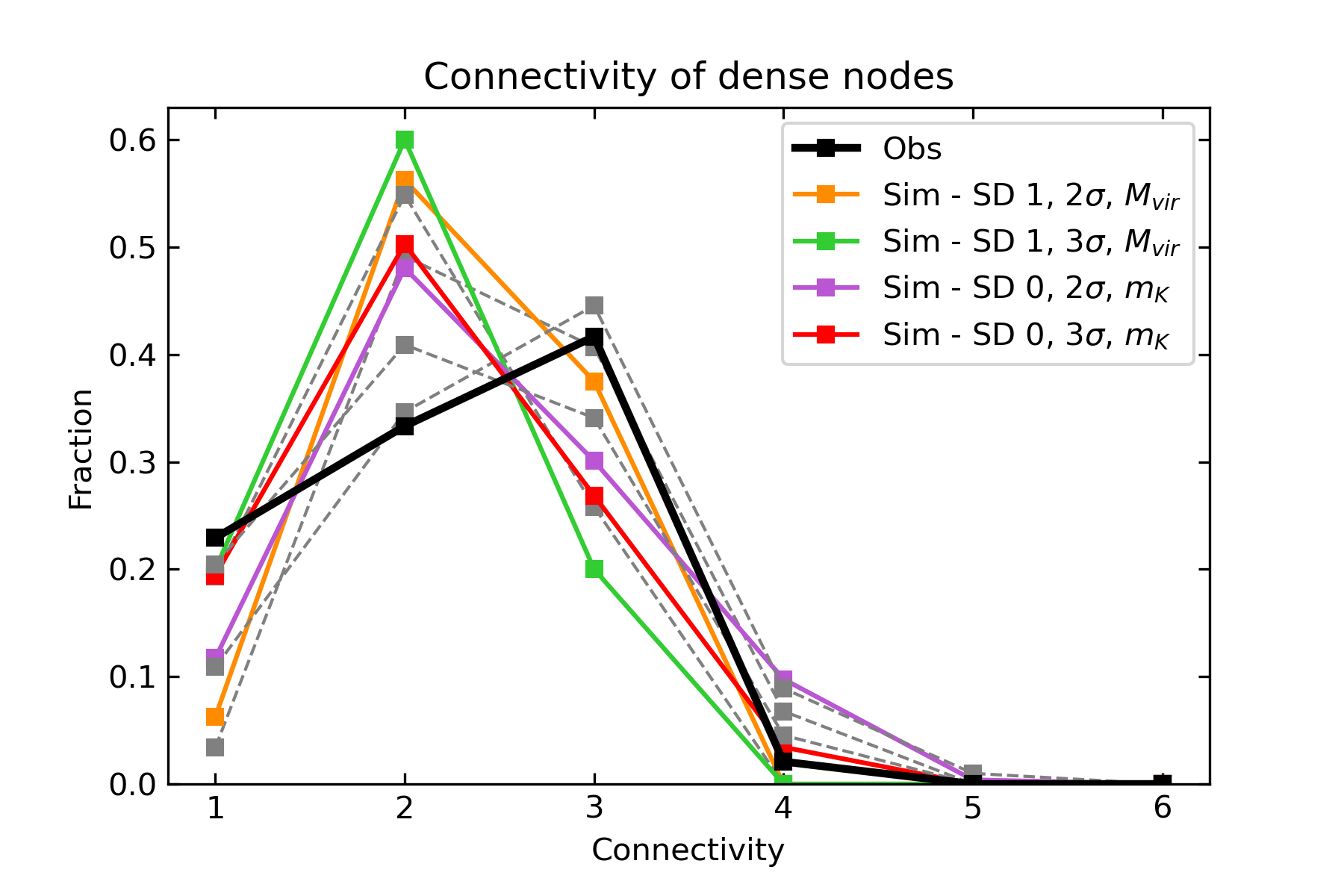}}
\caption{Node connectivity distribution. The black bold distribution refers to observed nodes from \citet{Malavasi2020Coma} with $\mathrm{SD1} - 3\sigma$. The colored lines refer to simulated nodes with parameter combinations $\mathrm{SD0} - 2\sigma$ and $\mathrm{SD0} - 3\sigma$ (for magnitude selected galaxies, in purple and red, respectively), and $\mathrm{SD1} - 2\sigma$ and $\mathrm{SD1} - 3\sigma$ (for mass selected galaxies, in orange and green, respectively). Dashed grey lines represent the rest of the parameter combinations and provide an idea of the range of connectivities explored. Only nodes within a slice of thickness $\pm 75 \mathrm{Mpc}$ (50.8 Mpc/$h$) from the location of the (simulated or observed) Coma cluster and with overdensity $\log (1+\delta) = \log (1+\delta_{\mathrm{Coma}}) \pm 0.5$ (see text) were used to derive the distributions.}
\label{conndist_densnodes}
\end{figure}

We then further refine our selection of the nodes for the comparison of statistical properties of the skeleton. First, we select only nodes in a slice of thickness $\pm 75 \mathrm{Mpc}$ (50.8 Mpc/$h$) from the location of the (simulated or observed) Coma cluster and with an overdensity of $\log (1+\delta) = \log (1+\delta_{\mathrm{Coma}}) \pm 0.5$, where $\log (1+\delta_{\mathrm{Coma}}) = 1.64$ is the average overdensity of the critical points found within the virial radius of the cluster in observations. We show the connectivity distribution for these nodes in Figure \ref{conndist_densnodes}. In this case, the peak at $\kappa = 3$ is less evident and the low connectivity values are more populated. We find that in this case, low persistence ($2\sigma$) parameter combinations are closer to observations.

\begin{figure}
\centering
\resizebox{\hsize}{!}{\includegraphics{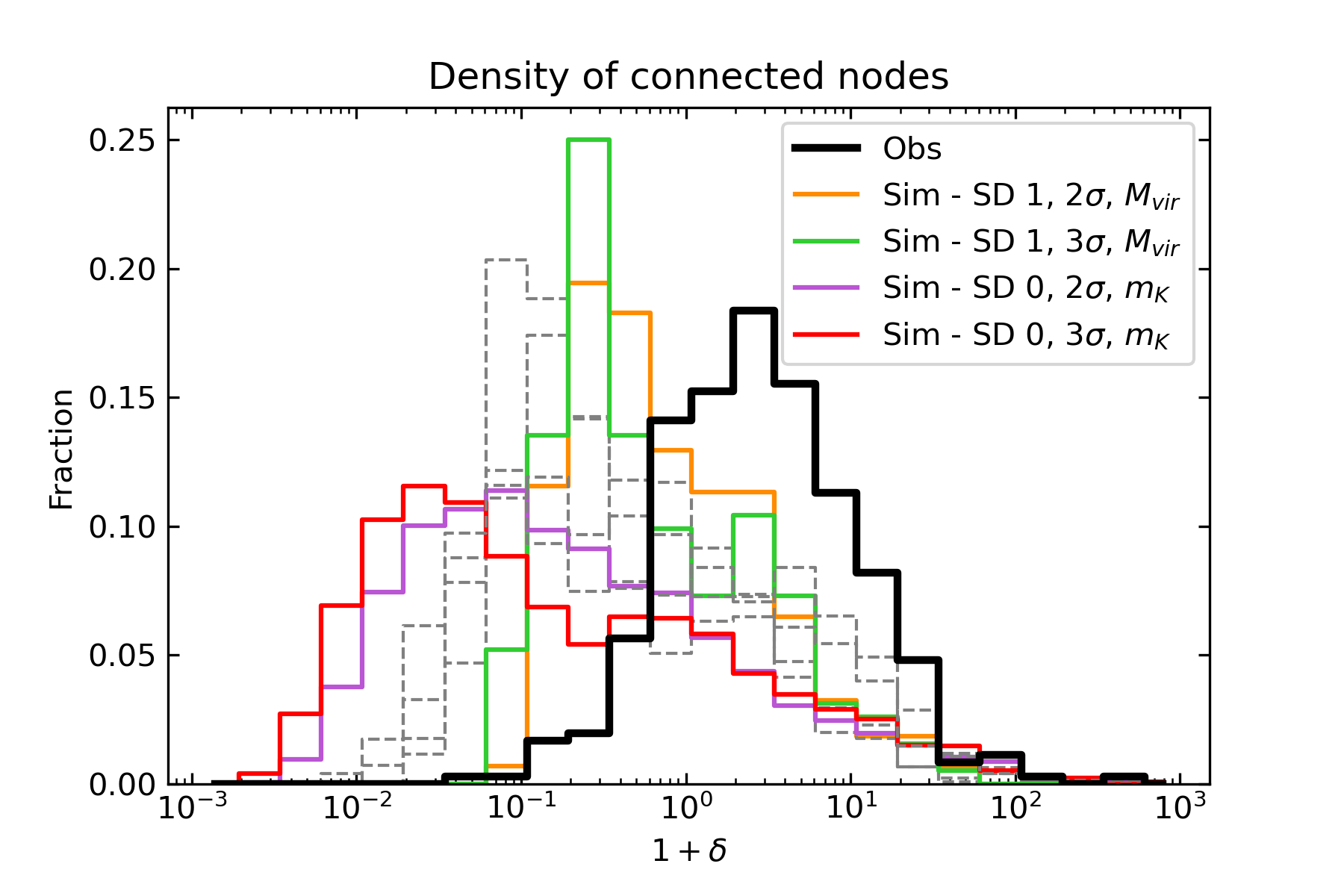}}
\caption{Node density distribution. The black bold distribution refers to observed nodes from \citet{Malavasi2020Coma} with $\mathrm{SD1} - 3\sigma$. The colored lines refer to simulated nodes with parameter combinations $\mathrm{SD0} - 2\sigma$ and $\mathrm{SD0} - 3\sigma$ (for magnitude selected galaxies, in purple and red, respectively), and $\mathrm{SD1} - 2\sigma$ and $\mathrm{SD1} - 3\sigma$ (for mass selected galaxies, in orange and green, respectively). Dashed grey lines represent the rest of the parameter combinations and provide an idea of the range of connectivities explored. Only nodes within a slice of thickness $\pm 75 \mathrm{Mpc}$ (50.8 Mpc/$h$) from the location of the (simulated or observed) Coma cluster and with $\kappa = 2 \div 3$ are used to derive the distributions.}
\label{densdist_connnodes}
\end{figure}

As a last test, we select nodes in a slice of thickness $\pm 75 \mathrm{Mpc}$ (50.8 Mpc/$h$) from the location of the (simulated or observed) Coma cluster and with a connectivity value of $\kappa = 2 \div 3$, i.e. in line with the connectivity of the Coma cluster from observations. We measure the $1+\delta$ distribution for maxima and bifurcations in this connectivity range, which we show in Figure \ref{densdist_connnodes}. This Figure shows that nodes in observations are distributed at higher densities than nodes in simulations. While no density distribution from the simulated skeletons is fully in the range of densities covered by observations, skeletons with large smoothing levels (SD1) are closer to the observed skeleton. In this case too, the highlighted parameter combinations are found at the extrema of the range of behaviors shown by the simulated skeletons.

\subsection{The cosmic web around Coma with the FoG effect}
\label{subsfog}
In Table \ref{cp_inside_coma_vpec} we report the version of Table \ref{cp_inside_coma} we obtain when we introduce FoG distortions in our galaxy population. We stress that the FoG distortions are introduced using all galaxies in the box, therefore for all clusters, not just Coma. This distorts the entire skeleton we recover in the box and not just the cosmic web around Coma. The fact that we still detect a realistic cosmic web connected to the Coma cluster is thus a success of our method. One of the differences between the cosmic web detected around Coma with and without FoG distortion is that we had to increase the radius of the sphere centered on Coma inside which we match critical points to the cluster. While this radius was set to $r_{\mathrm{vir}}$ for the case without FoG effect for both the mass selected and the magnitude selected sample, we resolve to set this search radius to $1.5 \times r_{\mathrm{vir}}$ (for the combination $\mathrm{SD0} - 2\sigma$), $2.5 \times r_{\mathrm{vir}}$ (for the combination $\mathrm{SD0} - 3\sigma$), and $3.5 \times r_{\mathrm{vir}}$ (for the combinations $\mathrm{SD1} - 2\sigma$ and $3\sigma$) for mass selected galaxies. This is due to the fact that no critical points can be found within one virial radius of our simulated Coma in the case of mass selected galaxies. The value of the search radius is still set to $r_{\mathrm{vir}}$ for magnitude selected galaxies\footnote{If instead of using $v_{\mathrm{pec}}$ computed from the output of our simulations we use the random extraction from Gaussians based on the computation of the velocity dispersion for each main halo, then the values of the search radius used to match critical points to Coma are $r_{\mathrm{vir}}$ for magnitude selected galaxies and $1.5 \times r_{\mathrm{vir}}$ for mass selected galaxies (all levels of smoothing of the density field and persistence thresholds).}.

\begin{table*}
\caption{Critical points associated with Coma. The first four columns refer to the critical points detected with the various combinations of \disperse~parameters within a few virial radii of the simulated Coma (see text and first row of each sub-table, where the exact number of virial radii for the search of critical points to be matched with the simulated Coma is expressed). The last column provides the critical points detected within the virial radius of the real Coma in \citet{Malavasi2020Coma}. In this case, FoG distortions are introduced in the galaxy distribution prior to the filament extraction with \disperse.}           
\label{cp_inside_coma_vpec}   
\centering                                   
\begin{tabular}{c c c c c | c}         
\hline\hline                       
\multicolumn{6}{c}{Mass selected galaxies} \\                         
\hline
CP type & $\mathrm{SD0} - 2\sigma$ & $\mathrm{SD0} - 3\sigma$ & $\mathrm{SD1} - 2\sigma$ & $\mathrm{SD1} - 3\sigma$ & \citet{Malavasi2020Coma} \\   
\hline
$n \times r_{\mathrm{vir}}$ & 1.5 & 2.5 & 3.5 & 3.5 & 1 \\
Minima 	   & 0 & 0 & 0 & 0 & 0 \\  
1-saddles    & 0 & 0 & 0 & 0 & 0 \\  
2-saddles    & 0 & 0 & 0 & 0 & 0 \\  
Maxima  	   & 0 & 0 & 0 & 0 & 1 \\ 
Bifurcations & 1 & 1 & 2 & 2 & 1 \\
\hline                                        
\multicolumn{6}{c}{Magnitude selected galaxies} \\                           
\hline
CP type & $\mathrm{SD0} - 2\sigma$ & $\mathrm{SD0} - 3\sigma$ & $\mathrm{SD1} - 2\sigma$ & $\mathrm{SD1} - 3\sigma$ & \citet{Malavasi2020Coma} \\   
\hline
$n \times r_{\mathrm{vir}}$ & 1 & 1 & 1 & 1 & 1 \\
Minima 	   & 0 & 0 & 0 & 0 & 0 \\  
1-saddles    & 0 & 0 & 0 & 0 & 0 \\  
2-saddles    & 0 & 0 & 0 & 0 & 0 \\  
Maxima  	   & 0 & 0 & 0 & 1 & 1 \\ 
Bifurcations & 1 & 1 & 1 & 1 & 1 \\
\hline
\end{tabular}
\end{table*}

\begin{figure*}
\sidecaption
\includegraphics[trim = 0.75cm 4cm 1cm 4.2cm, clip = true, width = 12cm]{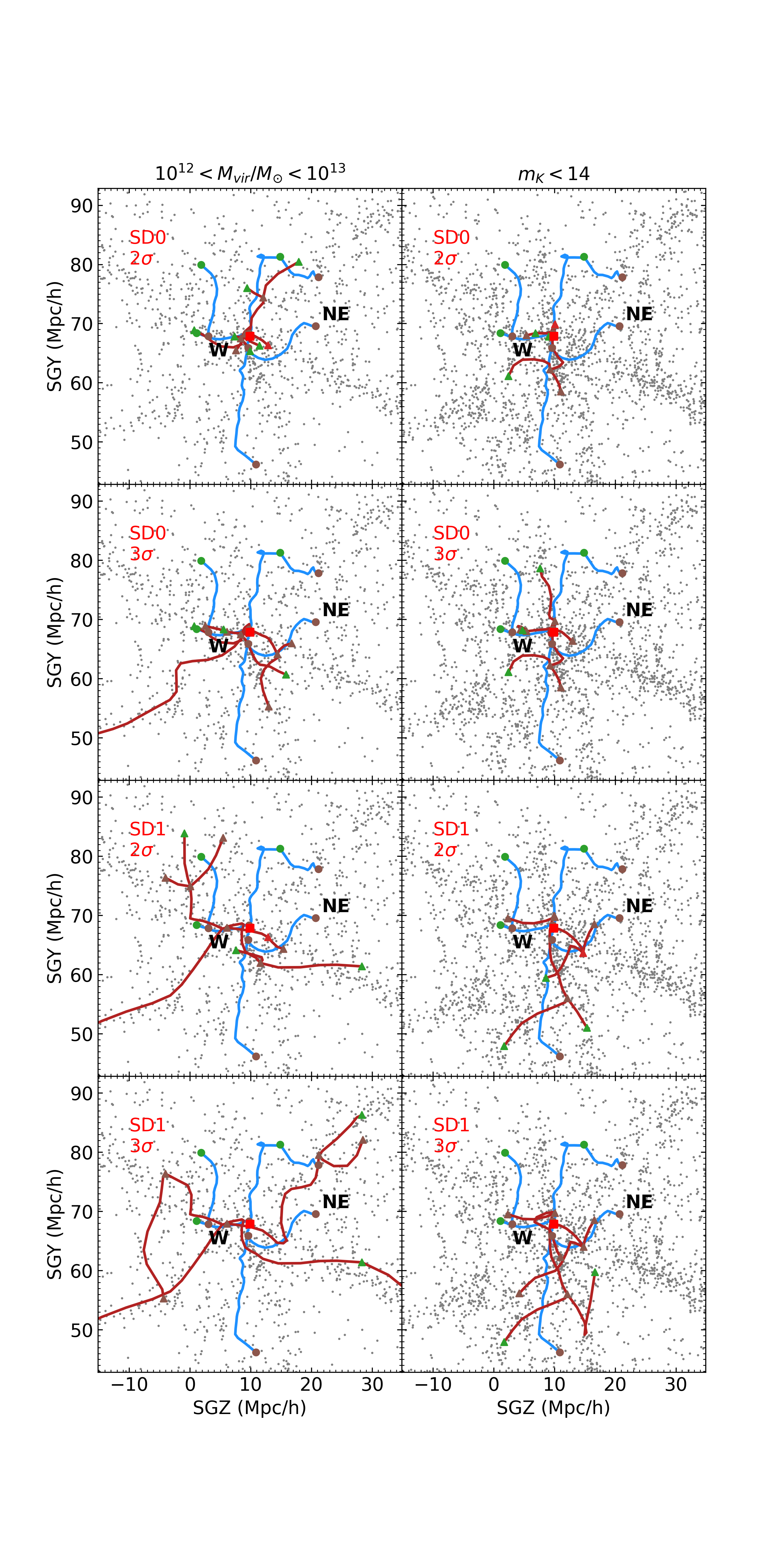}
\caption{Filaments in the constrained simulation compared to filaments in observations when the FoG effect is introduced in our simulation. The left column shows the case of a mass selected galaxy sample in the constrained simulation, the right column shows the case of a magnitude selected galaxy sample (see text). In each panel, simulated galaxies are shown as grey points in a slice of thickness 50 Mpc/h centered on Coma. Blue lines are filaments from \citet{Malavasi2020Coma} in supergalactic coordinates, the red square marks the position of the Coma cluster. Dark red lines are filaments obtained in the simulation with a variety of \disperse~parameter combinations, namely: $\mathrm{SD0} - 2\sigma$ (top row), $\mathrm{SD0} - 3\sigma$ (second row), $\mathrm{SD1} - 2\sigma$ (third row), $\mathrm{SD1} - 3\sigma$ (bottom row). The observed NE and W filaments are marked for reference. Circle and triangles mark the positions of critical points (in observations and simulations, respectively) and are color-coded according to their type (red: maxima, green: type 2 saddles, brown: bifurcations). This figure shows the SGZ-SGY projection.}
\label{visual_skel_comparison_proj1_vpec}
\end{figure*}

Figure \ref{visual_skel_comparison_proj1_vpec} shows the new skeleton detected in the vicinity of the Coma cluster when the FoG effect is included in the analysis of our simulation.  

This figure shows that the cosmic web we detect introducing FoGs is indeed realistic and close to the observed one. In the case of magnitude selected galaxies, for all smoothing levels and persistence threshold we recover a filament to the west and one to the north-east, more prominent for higher smoothing levels. The presence of these filaments is consistent with the case in which no FoGs are introduced. Moreover, filaments are now detected along the FoG direction (which were not present in the case without FoGs), confirming our assumption that this filament is due to the FoG distortion effect. 

In the case of mass selected galaxies, the situation is more complex. The large amount of virial radii that we need to reach in order to match critical points to the simulated Coma cluster indicates that the reconstruction of the cosmic web around the simulated Coma in this case is more uncertain. In the case of mass selected galaxies we recover a filament along the FoG direction for Coma only in the case $\mathrm{SD0} - 2\sigma$ and only hints of this in the other smoothing level and persistence combinations. In all combinations we do, however, recover both the west and the north-east filaments. Similar conclusions (albeit less evident) are reached also when FoGs are introduced using random extracted velocities for galaxies.

\section{Results}
In the previous sections we have proved that we can reconstruct a realistic simulated skeleton in the vicinity of the Coma cluster using our constrained simulation and we have tested that the reconstruction is robust against the fact that redshift space distortions are present in our observed data. In the following we will use the simulated skeleton derived without the introduction of the FoG effect to perform our analysis. We do so as we consider this skeleton to be more reliable, all the while we make sure that our results do not change if we use the skeleton with the inclusion of the FoG effect.

Based on our previous analysis we identify in the combination of \disperse~parameters $\mathrm{SD0} - 2\sigma$, galaxy selection based on apparent K-band magnitude and skeleton computed without introducing redshift space distortions our reference simulated filaments around Coma. In the following we will use this reference skeleton for visual representation, while we will include all \disperse~parameter combinations and galaxy selection criteria (namely $\mathrm{SD0} - 2\sigma$ and $\mathrm{SD0} - 3\sigma$ for magnitude selected galaxies and $\mathrm{SD1} - 2\sigma$ and $\mathrm{SD1} - 3\sigma$ for mass selected galaxies) in our statistical measurements.

In the remainder of the paper we will analyze the properties of the cosmic web connected to the simulated Coma cluster in comparison with observations. We will start by studying the connectivity of the cluster (i.e. the number of connected filaments) and then move on to the velocity distribution of the halos around the cluster and the filaments.

\subsection{The connectivity of the Coma cluster}
\label{simcomaconn}
In \citet{Malavasi2020Coma} we measured the connectivity of the real Coma cluster by counting the number of filaments crossing a sphere of radius $1.5 \times r_{\mathrm{vir}}$ centered on the cluster. We detected a connectivity $\kappa =  2 \div 3$ (median connectivity $\kappa = 2.5$). This is in line with the average connectivity of the nodes in both our simulated box and in observations (see e.g. Figures \ref{conndist} and \ref{conndist_densnodes}).

\begin{figure}
\centering
\resizebox{\hsize}{!}{\includegraphics{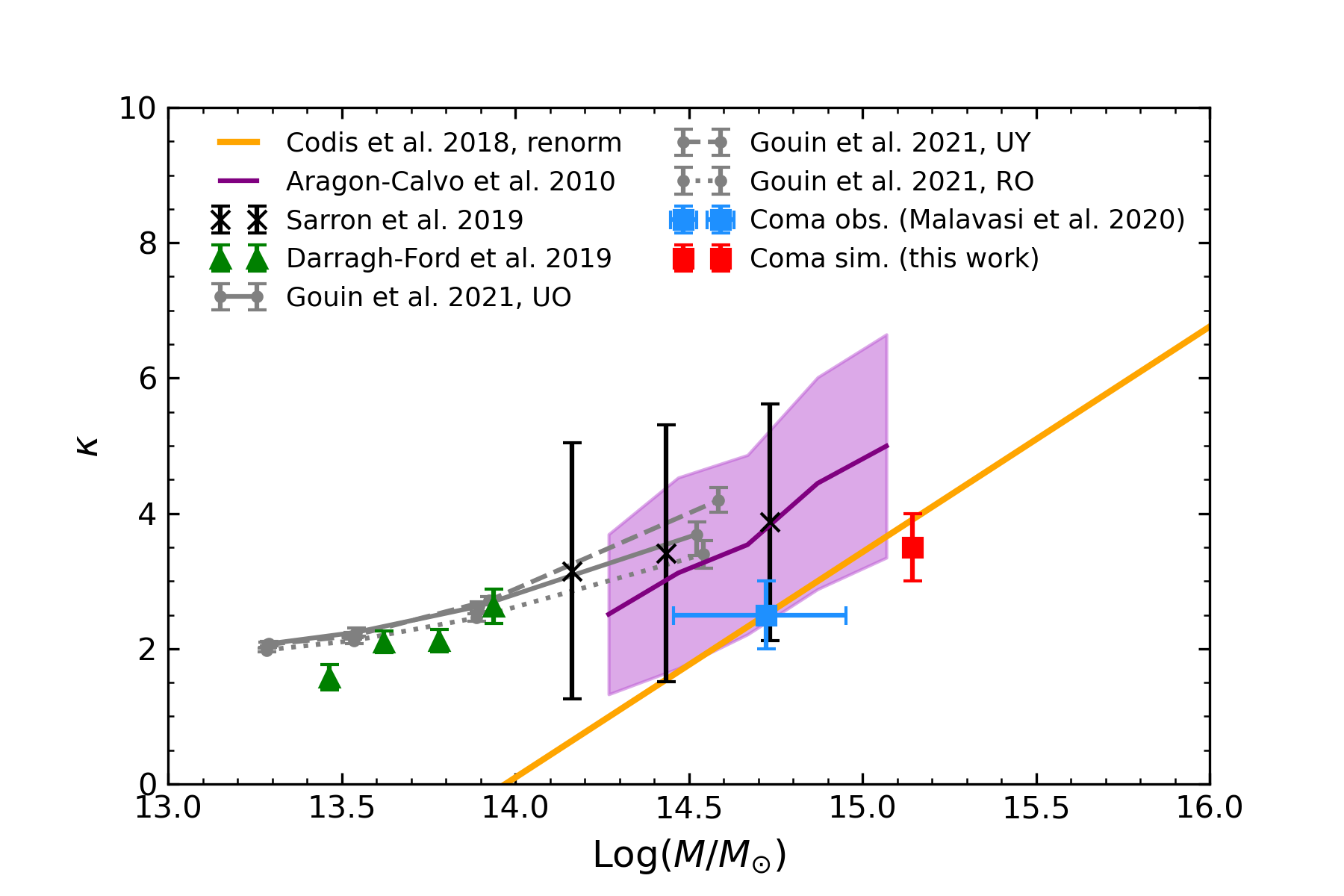}}
\caption{Connectivity of the real and simulated Coma clusters as a function of mass. The cyan square refers to the real Coma cluster as analyzed in \citet{Malavasi2020Coma}, while the red square corresponds to the simulated Coma cluster as detected in this work. Black crosses are the observed connectivity values from the AMASCFI Clusters \citep{Sarron2018, Sarron2019}, green triangles are the observed connectivity values for the groups in COSMOS \citep{DarraghFord2019}. Grey lines and points are connectivity measurements derived in the IllustrisTNG simulation \citep{Gouin2021} for clusters and groups which are unrelaxed and old (UO, solid line), relaxed and old (RO, dotted line), and unrelaxed and young (UY, dashed line). The purple line and shaded region are the relation from the numerical N-body simulations of \citet{AragonCalvo2010}, and its corresponding $1\sigma$ uncertainty, while the solid orange line is the theoretical relation of \citet{Codis2018} renormalized to pass through our measurement for the real Coma cluster, so as to provide a better comparison of trends. We note that the mass on the $x$-axis of the plot is $M_{200}$ for the real and simulated Coma, the measurements by \citet{DarraghFord2019}, and those of \citet{Sarron2019}, while it is a value close to $M_{\mathrm{vir}}$ for the \citet{AragonCalvo2010} relation.}
\label{comaconnectivity}
\end{figure}

We have measured the connectivity for the simulated Coma in the same way. Starting from the list of critical points associated with the cluster (Table \ref{cp_inside_coma}), we have counted all the filaments with one of those critical points as an extreme and the other one located outside a sphere of $1.5 \times r_{\mathrm{vir}}$. We report a median connectivity for simulated Coma of $\kappa = 3.0$ (with the connectivity in the range $\kappa = 3 \div 4$ depending on the \disperse~parameter combination and galaxy selection criterion). This is in good agreement with our observations, with the slightly larger number of connected filaments possibly due to the larger amount of tracers and better definition of the filaments in the constrained simulation. In Figure \ref{comaconnectivity} we show a rendition of Figure 5 of \citet{Malavasi2020Coma}, which shows the connectivity of the real Coma, simulated Coma, and other cluster samples from observations and simulations as a function of mass. The other theoretical and observed samples of clusters for which the connectivity has been measured are the AMASCFI clusters (\citealt{Sarron2018}, with the connectivity measured by \citealt{Sarron2019}), the groups detected in COSMOS by \citet{DarraghFord2019}, the connectivity-mass relation identified in N-body simulations by \citet{AragonCalvo2010} and the theoretical one derived by \citet{Codis2018} renormalized to pass through the point corresponding to the real Coma as done in \citet{Malavasi2020Coma}. To these we have added the connectivity-mass trends derived by \citet{Gouin2021} in the IllustrisTNG simulation \citep{Naiman2018, Marinacci2018, Springel2018, Nelson2018, Pillepich2018}, for relaxed old clusters (RO), and unrelaxed old and young ones (UO and UY, respectively, see \citealt{Gouin2021} for the distinction between relaxed and unrelaxed, old and young clusters). As for the measurements related to the real and simulated Coma, they are placed on this plot based on the measured median connectivity (with error bars corresponding to the measured connectivity range) and according to their $M_{200}$ value (measured by \citealt{Gavazzi2009} in the case of the real Coma). We have converted the $M_{200}$ value for the simulated Coma to the same cosmology we adopted in \citet{Malavasi2020Coma}. We report the error bars on the measurement for $M_{200}$ in the case of the real Coma to show that, although the measurement of the mass is lower than what we find in the simulation, there could still be room for agreement thanks to the uncertainty on the observed value (see also the discussion in Section \ref{identifyingcoma}).

Even if the caveats on the connectivity measurements as a function of mass for observed and simulated clusters from the literature still apply to this situation\footnote{We refer the reader to Section 4.2 of \citet{Malavasi2020Coma} for the full discussion which here we only briefly summarize: the connectivity for the AMASCFI clusters was derived in a way different from our measurement, i.e. by counting filaments which cross a sphere of 1.5 Mpc (1.02 Mpc/$h$) centered on each cluster, which in our case would be completely within the virial radius sphere of the real and simulated Coma. Moreover the connectivity-mass relation for the N-body simulations of \citet{AragonCalvo2010} is reported using a mass value more similar to $M_{\mathrm{vir}}$ than to $M_{200}$. In the case of \citet{DarraghFord2019} and \citet{Gouin2021}, connectivity is measured in a way similar to what we do with Coma, i.e. by counting the number of filaments that cross a sphere of radius $1.5 \times r_{\mathrm{vir}}$ or $1.5 \times R_{200}$, respectively.}, we find agreement between the connectivity value at the mass of the simulated Coma and the values from the literature reported in this plot. The connectivity we measured for the simulated Coma is at the lower edge of the general trend identified by the literature values, indicating moderate agreement or a slightly lower value. As this is the same conclusion we found for the real Coma, investigating why these massive clusters have lower connectivity values than what expected given their mass, whether this is a real effect or due to the way in which connectivity is measured, and how this picture evolves with redshift strongly calls for a deeper investigation of the connectivity-relation from low to high redshift and from the group to the cluster (and potentially super-cluster) regime, which will be the subject of a future work.

\subsection{Accretion on the Coma cluster through the cosmic web}
\label{accretion}
We explore the dynamics of matter in the vicinity of the simulated Coma cluster and its connected filaments. To do this we make use of the velocities of the dark matter halos in the simulation box. Each halo is characterized by a velocity vector with components $\boldsymbol{v} = (v_x, v_y, v_z)$. For each halo (including those selected as galaxies), we subtract the velocity vector of the halo identified as the simulated Coma: $\boldsymbol{v}_{h} = \boldsymbol{v} - \boldsymbol{v}_{\mathrm{Coma}}$. We then compute the dot product of the velocity for each halo and the difference in position between each halo and the position of the simulated Coma ($\boldsymbol{r}_{h} = \boldsymbol{r} - \boldsymbol{r}_{\mathrm{Coma}}$). The dot product

\begin{equation}\label{vradmoduledef}
v_{\mathrm{rad}, \mathrm{C}} = \frac{\boldsymbol{v}_{h} \cdot \boldsymbol{r}_{h}}{r_h}
\end{equation}

where $r_h$ is the norm of the vector $\boldsymbol{r}_{h}$, provides the module of the velocity of each halo, projected along the line connecting the simulated Coma cluster with that halo, in a reference frame where Coma is at the origin and at rest. We then construct the 3D components of such an array in the form:

\begin{equation}\label{vradcomponentsdef}
\begin{cases}
v_{\mathrm{rad}, \mathrm{C}, x} = v_{\mathrm{rad}, \mathrm{C}} \times r_{h, x}/r_h \\
v_{\mathrm{rad}, \mathrm{C}, y} = v_{\mathrm{rad}, \mathrm{C}} \times r_{h, y}/r_h \\
v_{\mathrm{rad}, \mathrm{C}, z} = v_{\mathrm{rad}, \mathrm{C}} \times r_{h, z}/r_h
\end{cases}
\end{equation}

\begin{figure*}
\centering
\resizebox{\hsize}{!}{\includegraphics[trim = 1.5cm 1.5cm 1.5cm 1.5cm, clip = true]{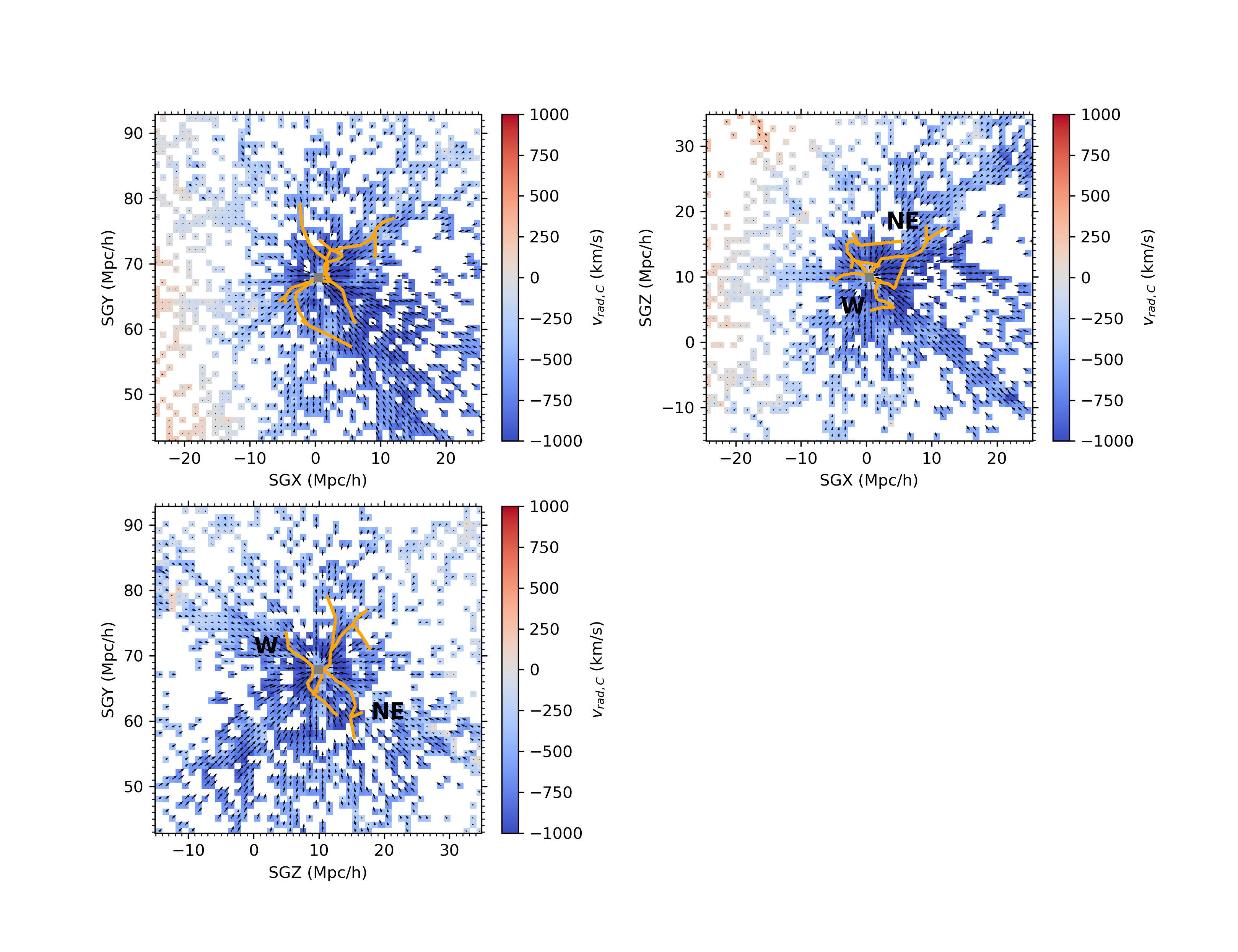}}
\caption{Maps of halo velocities around the simulated Coma cluster. The grey square at the center marks the position of the Coma cluster. Orange lines are first, second, and third generation filaments (the case $\mathrm{SD0} - 2\sigma$ for magnitude selected galaxies is reported here as an example, see brown lines in Figure \ref{visual_skel_comparison_proj1} and text). The top left panel shows the $x$-$y$ projection, the top right panel shows the $x$-$z$ projection, and the bottom left panel shows the $z$-$y$ projection (shown in the first panel on the right column of Figure \ref{visual_skel_comparison_proj1}). In each slice, the color map shows a 2D histogram in which in each bin the average $v_{\mathrm{rad}, \mathrm{C}}$ is measured (red stands for halos moving away from Coma, blue for halos moving towards Coma, white pixels stand for empty bins). Black arrows are constructed from the average in each bin of the $v_{\mathrm{rad}, \mathrm{C}}$ components relative to the projection shown (i.e. $v_{\mathrm{rad}, \mathrm{C}, x}$ and $v_{\mathrm{rad}, \mathrm{C}, y}$ in the top left panel, $v_{\mathrm{rad}, \mathrm{C}, x}$ and $v_{\mathrm{rad}, \mathrm{C}, z}$ in the top right panel, $v_{\mathrm{rad}, \mathrm{C}, z}$ and $v_{\mathrm{rad}, \mathrm{C}, y}$ in the bottom left panel). The thickness of the considered slice is of 20 Mpc/h, all halos in the slice are considered, regardless of mass or whether they are selected as galaxies. The labels mark the position of the simulated NE and W filaments.}
\label{halovelocities}
\end{figure*}

We show maps of these quantities in relation to the filament positions in the surroundings of Coma in Figure \ref{halovelocities}. This figure shows three projections of the 3D system formed by Coma and its surrounding filaments. In each projection, the color map shows a 2D histogram in which in each bin the average $v_{\mathrm{rad}, \mathrm{C}}$ is computed. Black arrows, instead, show the average direction computed only from the components relevant to the considered slice (e.g. in the $x$-$y$ projection each black arrow is constructed from the average in the 2D histogram bin of $v_{\mathrm{rad}, \mathrm{C}, x}$ and $v_{\mathrm{rad}, \mathrm{C}, y}$). This gives an idea of the module and direction of the velocity of the halos in the simulation box in proximity of the simulated Coma and its filaments. Only the case of magnitude selected galaxies and $\mathrm{SD0} - 2\sigma$ is shown for reference in this figure, while all four parameter combinations described above as the closest to observations are considered when the infall on Coma from the filaments is quantified in the following.

This figure shows how there is a significant number of halos moving towards Coma from the large scale structure. The velocity of the halos seems to correlate fairly well with the filament position. This is particularly evident e.g. in the $x$-$y$ and $z$-$y$ projections, especially for the filaments to the north east and the west of the cluster in the latter.

\begin{figure*}
\sidecaption
\includegraphics[trim = 0.5cm 1.5cm 1.5cm 1.5cm, clip = true, width = 12cm]{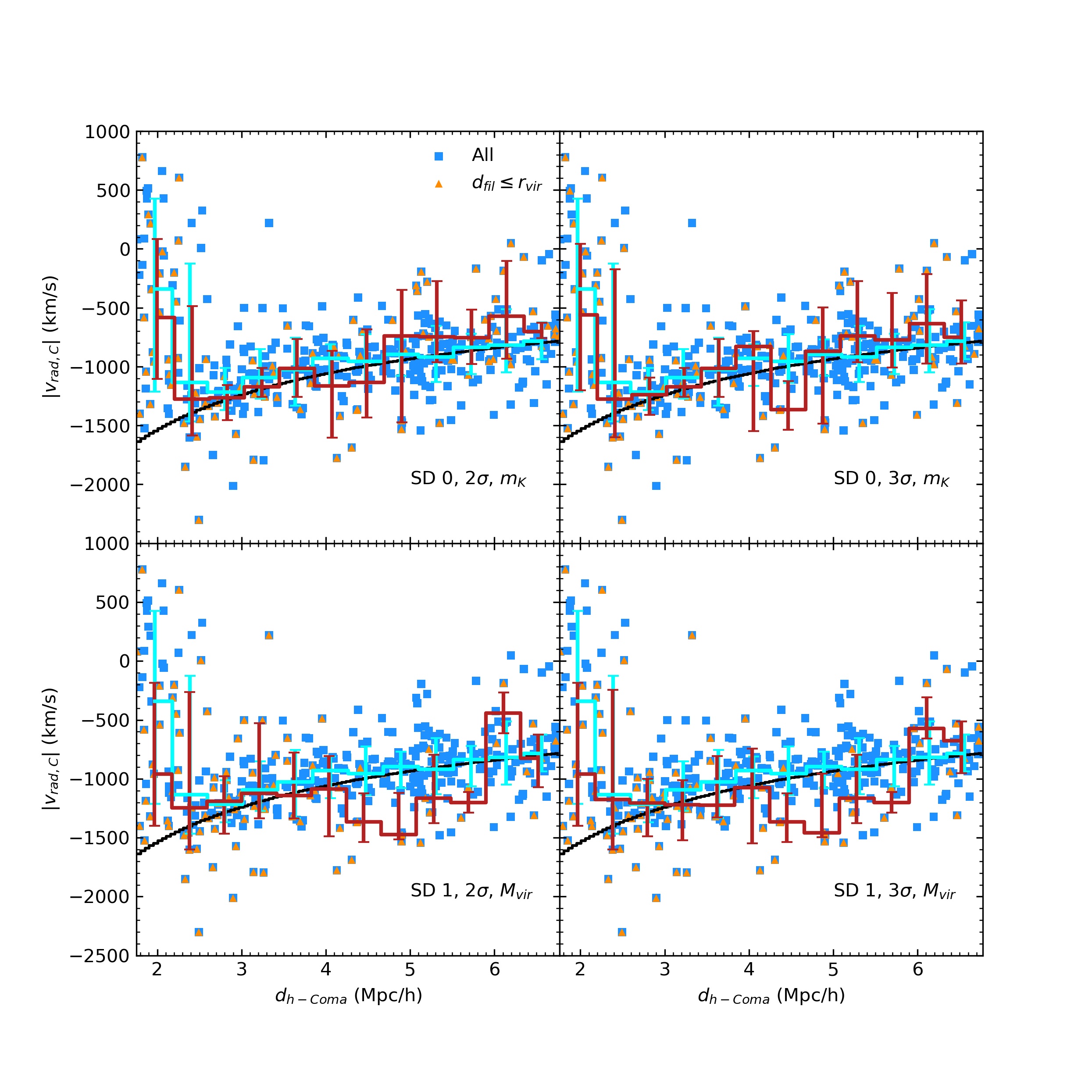}
\caption{Halo velocity as a function of distance from Coma. Each panel shows the halo velocity $v_{\mathrm{rad}, \mathrm{C}}$ as a function of the distance between each halo and the center of the simulated Coma cluster $d_{h-\mathrm{Coma}} = |\boldsymbol{r}_{h}|$. Only halos with $r_{\mathrm{vir}} \leq d_{h-\mathrm{Coma}} \leq 10 Mpc$ (6.774 Mpc/$h$) have been considered. Blue squares refer to all halos within this distance range, orange triangles are halos with $d_{\mathrm{fil}}$ smaller than the 25th percentile of the $d_{\mathrm{fil}}$ distribution. Solid lines refer to the running means of the distributions (cyan: all halos, red: halos close to filaments), error bars encompass the region between the 16th and the 84th percentile of the distributions. Different panels refer to different combinations of \disperse~parameters and galaxy selections. In each panel, the black line shows the velocity that a body would reach as a function of distance from Coma by starting at 10 Mpc (6.774 Mpc/$h$), with a velocity equal to the value of the last bin of the running mean as derived for the total halo population, and being subject only to the gravitational attraction of the Coma cluster.}
\label{phasespace}
\end{figure*}

We try to quantify the relation between halo velocity (whether infalling or outflowing) and filament direction. In Figure \ref{phasespace} we report the value of $v_{\mathrm{rad}, \mathrm{C}}$ as a function of halo distance from Coma. We consider all the halos which are outside of the simulated Coma virial radius (so as to reduce the influence of the cluster on the galaxy motions) and within 10 Mpc (6.774 Mpc/$h$) from the cluster center. We chose this limit as we see from Figure \ref{halovelocities} that even third generation filaments do not generally extend beyond such radius. The general distribution of halos in this region shows that infall velocity tends to increase with decreasing distance from the simulated Coma center, except in the close vicinity of the cluster, where galaxies tend to have a larger scatter around zero, with some having large outflow velocities. This could be due to galaxies that have already gone through the cluster once and are coming out the other side of the cluster with a positive velocity or to galaxies that are close to being virialized and that are within the zero-velocity surface of the cluster. At distances from Coma larger than 2.5 Mpc/h, velocities are generally negative, implying infall on the cluster. In the same figure, we also show the velocity that a body would reach as a function of distance from Coma by starting at 10 Mpc (6.774 Mpc/$h$), with a velocity equal to the value of the last bin of the running mean as derived for the total halo population, and being subject only to the gravitational attraction of the Coma cluster. The velocity distribution for the total halo population closely follows this relation (except close to the virial radius of the cluster), implying that no additional effect besides the overall gravitational attraction from the cluster is at play here. We derive the relation between infall velocity and distance from the simulated cluster also only for those halos that are closer to the axes of the filaments than a certain threshold. For each halo, we measure the distance between it and the axis of the closest filament ($d_{\mathrm{fil}}$) by using the common formula for the distance between a point and a line (where the line in this situation is the closest segment of the closest filament to the halo, using the \disperse~formalism)\footnote{We note that this distance definition is different than the one used e.g. in \citet[][see the definition of $d_{\mathrm{fil}}$]{Malavasi2022}. In the case of \citet{Malavasi2022}, $d_{\mathrm{fil}}$ was measured as the distance between a halo and the midpoint of the closest segment of the closest filament. The two distances are comparable in the case where filament segments are very short or if the halo is far away from a filament segment. In any case the updated version of $d_{\mathrm{fil}}$ we use here affects only a minority of halos and in a small way. We refer the reader to Figure 1 of \citet{Malavasi2022} for a schematic representation of $d_{\mathrm{fil}}$.}. As a threshold to consider a halo close enough from the axis of a filament to be reported on Figure \ref{phasespace}, we chose the 25th percentile of the $d_{\mathrm{fil}}$ distribution.

There is no obvious difference in the velocity distribution as a function of halo distance from Coma for halos in filaments compared to the general halo population. In both the case of filaments extracted with magnitude-selected and mass-selected galaxies, even the averages of the populations are largely overlapping, showing no difference. Only in the case of mass selected galaxies, a few of the bins located at distances from Coma of $\sim 4 \div 6$ Mpc show more negative values, suggesting larger inflow velocities. However, from this figure we are bound to conclude that the amplitude of the velocity of halos is somewhat uncorrelated from their position with respect to filaments, and it is mostly sensitive to the large-scale inflow of matter on the Coma cluster.

We therefore explored the relation between the distance of the halos from the axes of the filaments $d_{\mathrm{fil}}$ and the angle between the direction of the halo projected velocity ($\boldsymbol{v}_{\mathrm{rad}, \mathrm{C}} = (v_{\mathrm{rad}, \mathrm{C}, x}, v_{\mathrm{rad}, \mathrm{C}, y}, v_{\mathrm{rad}, \mathrm{C}, z}$) and the local direction of the closest filament ($a_{\mathrm{fil}}$). The local direction of the closest filament to a given halo is consistently chosen to be positive when pointing away from the Coma cluster, so as to be consistent with our definition of the projected velocity. In this reference frame, halos with velocities infalling towards Coma will have angles closer to $180 \deg$, while halos outflowing from the cluster would have an angle closer to $0 \deg$.

\begin{figure*}
\sidecaption
\includegraphics[trim = 0.5cm 1.5cm 1.5cm 1.5cm, clip = true, width = 12cm]{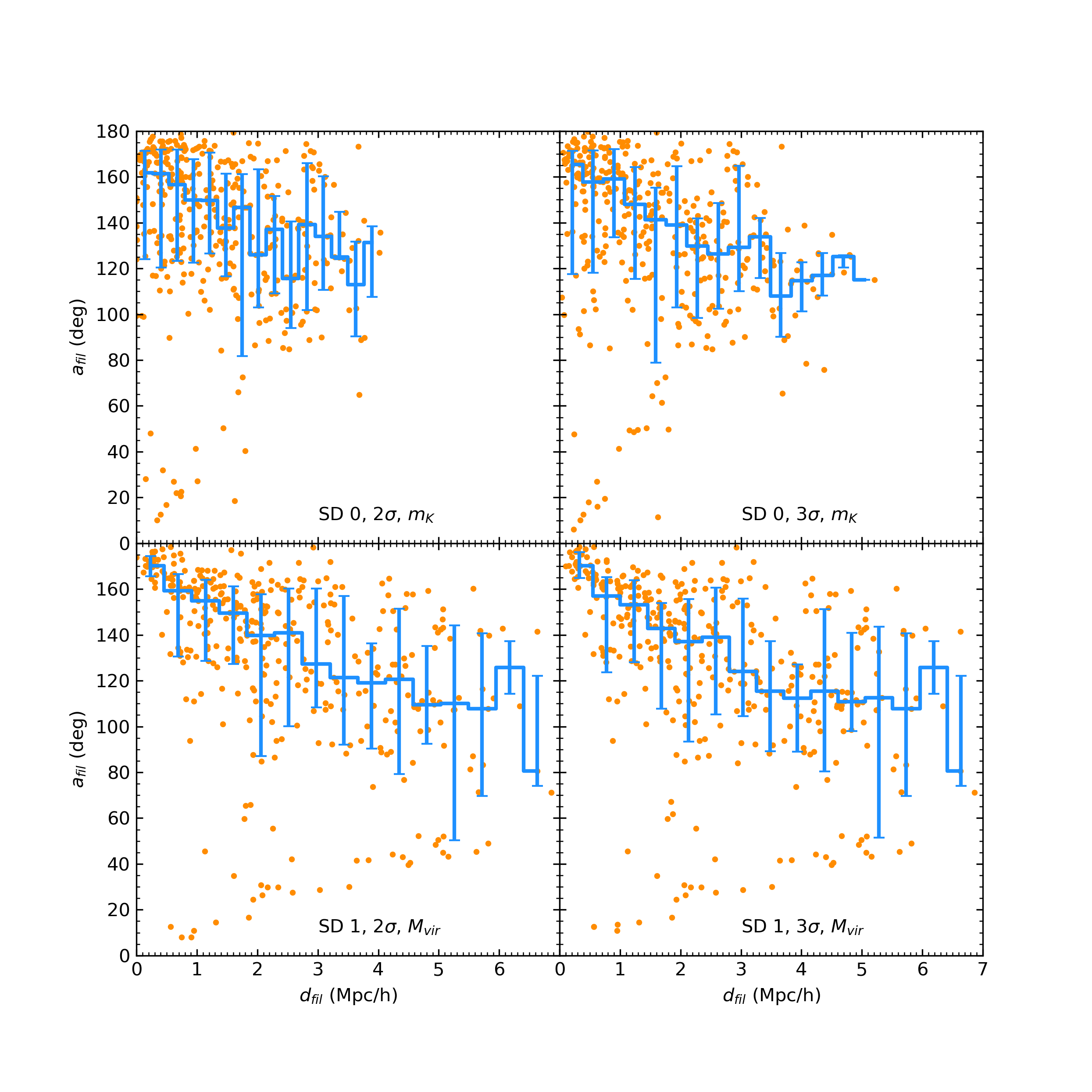}
\caption{Angle $a_{\mathrm{fil}}$ between the projected velocity of a halo ($\boldsymbol{v}_{\mathrm{rad}, \mathrm{C}}$) and the local direction of the closest filament as a function of the halo distance from the axis of the filaments $d_{\mathrm{fil}}$. Points represent the actual measurements, solid lines represent the running means of $a_{\mathrm{fil}}$ as a function of $d_{\mathrm{fil}}$, error bars encompass the region between the 16th and the 84th percentile of the distributions. Different panels refer to different combinations of \disperse~parameters and galaxy selections.}
\label{veldfil}
\end{figure*}

Figure \ref{veldfil} shows the relation between $a_{\mathrm{fil}}$ and $d_{\mathrm{fil}}$. The majority of the halos have angles close to $180 \deg$, as it is expected given the significant infall of material on the cluster. However, there is also a trend visible of $a_{\mathrm{fil}}$ becoming closer to $180 \deg$ with decreasing $d_{\mathrm{fil}}$. This provides an indication of the fact that the flux of matter infalling on Coma becomes more collimated around the location of the filaments. This is the first time that such a result is reported. Moreover, while this effect is small at low redshift, we expect it to become stronger at higher redshift when the accretion on the cluster from the filaments was higher. We will investigate this in a future work in this series.
 
We did perform the same analysis (in particular the one shown in Figures \ref{phasespace} and \ref{veldfil}) also on the cosmic web around our simulated Coma extracted when the FoG effect is included (see Section \ref{foginclusion}). In this case too, we do not see a difference in the velocity distribution as a function of halo distance from Coma for halos in filaments compared to the general halo population, which allows us to reach the conclusion that halos in filaments are not faster in terms of accretion velocity than the general halo population. When we analyze the $a_{\mathrm{fil}} - d_{\mathrm{fil}}$ relation, we do not recover trends of $a_{\mathrm{fil}}$ to become closer to $180 \deg$ with decreasing distance from the axes of the filaments, except for the case of magnitude selected galaxies and lower persistence threshold. However the value of $a_{\mathrm{fil}}$ remains stable at a large value close to $120 \div 140 \deg$ also at large values of $d_{\mathrm{fil}}$. This indicates that the flux of matter is collimated close to the filaments also when the FoG effect is included in the extraction of the skeleton.
 
\section{Discussion}
\label{discussion}
The first challenge we face in this paper, is determining how similar the filaments close to the simulated Coma are to the filaments around the real Coma. Our approach to this problem is different from what usually done in the literature due to the peculiar nature of our constrained simulations. For example, one approach commonly adopted is to apply different algorithms to detect the filaments to the same observed or simulated data set. \citet{Libeskind2018}, \citet{Rost2020}, and \citet{Bonnaire2020} perform this kind of comparisons by means of distributions of global quantities for the filaments (e.g. length, radius, luminosity, mass, volume, redshift) or, in the case of simulations, by also comparing which components are associated to the same structure by different algorithms. We adopted a similar approach in \citet{Malavasi2020Catalogue}, where we compared filament catalogues obtained from the same galaxy survey with the same algorithm, only run with varying parameters. In the case of the present work, this approach is less constraining. In fact, several of the filament properties (e.g. their length) could be influenced by the volume density of the tracers (i.e. galaxies), which is different in simulations and observations.

A complementary approach is to run the same algorithm on simulated samples of galaxies extracted from simulations light-cones tuned to reproduce a given galaxy survey, where the mock samples are slightly modified each time to test a different source of uncertainty in the filament reconstruction. In this case, a close correspondence between the various filament catalogues can be expected. More sophisticated approaches can then be undertaken to compare the catalogues, such as it was done in \citet{Malavasi2017, Laigle2018, Kraljic2018} where the distances of all filaments from one skeleton realization to the filaments of another one can be derived and compared. Also this approach is not viable in our case as, although our simulations are constrained to reproduce existing structures, the residual cosmic variance does not ensure a perfect correspondence.

We have therefore to resort mainly to visual comparison of the samples of filaments and to set for a match between observations and simulations in terms of number and spatial distribution of the filaments around the simulated cluster. We do conclude that our filament reconstruction is overall satisfactory. In the majority of cases we identify prominent NE and W filaments, connecting to the Coma cluster. This is important, as those are the filaments that observational evidence indicate as those being the most linked to features of the Coma cluster connected to matter accretion.

When investigating the dynamics of matter around the simulated Coma cluster and all the filaments in its vicinity, we found that there is limited evidence for the halo velocity to be higher in the proximity of the filaments, but stronger trends for the angle between the direction of the matter flow and the direction of the filament to be smaller. This points us towards the conclusion that the matter flow close to filaments is more collimated. This is the first work to report such a result in the literature.

This conclusion is in agreement with what found in several works in the literature. In particular, e.g. \citet{Dekel2009} highlighted the presence of high-flux, low-entropy streams accreting cold gas on a forming halo ($M_{\mathrm{vir}} > 10^{12} M_{\sun}$) at $z = 2.5$. Although this is a different mass and redshift range than what we explored in our paper, Figure 1 of \citet{Dekel2009} shows that while the velocity field magnitude does not seem to be different inside and outside the filaments, the flux is higher and the entropy lower inside the streams. This is in agreement with \citet{Danovich2012} who analyzed 350 halos of $10^{12} M_{\sun}$ at $z = 2.5$, finding that the streams of accreting gas are narrow and covering a small portion of the virial shell of the accreting halos. This is also in agreement with what found by \citet{BennettSijacki2020} for a $10^{12} M_{\sun}$ halo at $z = 6$, where they identify filaments of cold gas piercing through the shock at the virial radius of the halo. Although the radial velocity magnitude of the gas in the filaments is not much lower than the velocity of other accreting gas from the large scale, its turbulent velocity component is low, meaning that these stream of gas are collimated. This is supported also by the velocity dispersion of the gas in the filaments being lower than in the surrounding large scale structure (see their Figure 8). Cold gas filaments that accrete matter on the forming halo inside the virial radius are also found by \citet{Valentini2021}, for a $10^{12} M_{\sun}$ halo at $z = 6$. Coherent streams that accrete matter onto forming halos at high redshift are also highlighted in the simulations by \citet[][see their Figure 1]{AragonCalvo2019}, who also introduces the idea that when halos are detached from such flows galaxies can quench their star formation (see also \citealt{Moutard2022} where this effect, called "cosmic web detachment" is introduced as an explanation for the larger fraction of obscured, X-ray emitting AGNs in quenching galaxies and the change in spin direction of galaxies flowing along the cosmic web). Such accretion flows have also been found in observations. For example \citet{Martin2016} reports a filament with a collimated flow having a very low velocity dispersion ($\sim 50$~km/s) feeding a proto-disk in a $4 \times 10^{12} M_{\sun}$ halo at $z = 2.843$.

We stress the fact that the accretion flows of gas and dark matter are expected to be different, a fact that our N-body simulations are unable to probe. In fact, our analysis of the dark matter dynamics around Coma is based on the velocity of halos. Moreover, accretion of cold gas onto forming halos through filaments (a process explored in the literature cited above which uses hydrodynamical simulations) is a phenomenon limited to the high redshift Universe. Indeed, cold mode accretion is not the main mode of accretion onto halos today (clusters at $z = 0$ are expected to accrete mostly hot gas, inefficiently, through isotropic accretion more than from filaments of the cosmic web, see e.g. Figure 11 of \citealt{Overzierreview} and \citealt{DekelBirnboim2006}). What we detect around Coma may be only minor accretion from filaments still happening at $z = 0$. However, we also conclude that our findings provide a compelling science case for the investigation of the evolution of the cosmic web around this simulated Coma cluster and how the accretion of matter from the filaments changed as a function of redshift and of the different configurations of the filaments themselves (e.g. because of the possible merging of the filaments, \citealt{Cadiou2020}). This will be the subject of upcoming papers in this series.

\section{Conclusions}
\label{conclusion}
In this work, we explore the cosmic web around a reproduction of the Coma cluster of galaxies, identified in a box of $500\:\mathrm{Mpc}/h$ on a side simulated with a dark matter only run of the code \textsc{Ramses}. The initial conditions of this simulation have been constrained using the velocity of galaxies so that they reproduce existing structures in the local Universe.

We identify the halo corresponding to the Coma cluster in our simulation and we detect filaments connected to it with the \disperse~algorithm. We then analyze the filaments connected to the simulated cluster, comparing their number and spatial distribution with the observations of \citet{Malavasi2020Coma}, and the velocity distribution of the halos around these filaments to study the accretion of matter onto the cluster. We summarize our findings in the following way:

\begin{enumerate}
\item We report a good agreement between the cosmic web detected around the simulated Coma cluster and the large scale structure configuration in observations. We identify several combinations of the \disperse~parameter choices and galaxy selection criteria where prominent NE and W filaments are detected at the correct observed position.

\item The connectivity around the simulated Coma is in line with the value for the real Coma. However, given the difference between the mass of the real and simulated Coma (higher for the latter), the placement of the simulated Coma on the mass-connectivity relation is more uncertain. The value of $\kappa = 3$ is only in partial agreement with other values from the literature, both from observational and simulated samples.

\item The magnitude of the radial infall velocity components of the halos in a $10 \mathrm{Mpc}$ (6.774 Mpc/$h$) radius around the cluster shows no significant difference between the general halo population and a selection of only those halos which are close to the axis of the surrounding filaments. However, the angle between the radial infall velocity component vector and the local direction of the closes filament to a given halo become increasingly aligned with decreasing distance of the halos from the filaments axis. We conclude that the filaments convey a collimated flow of matter onto the simulated Coma cluster. This is the first time that such a result is reported for dark matter at low redshift.
\end{enumerate}

This paper is the first of a series, in which we aim to establish a correspondence between our constrained simulation and the cosmic web around the real Coma cluster. The additional power of this simulation is in its capability to offer us an evolutionary perspective on the Coma system. In future papers we will explore the evolution of the accretion of matter onto the Coma cluster from the filaments with redshift and how it impacted on the cluster formation.

\begin{acknowledgements}
%Referee
The authors would like to praise and thank the anonymous referee, whose on-point and in-depth comments helped greatly improve the quality of the paper.

%ByoPiC
Part of this research has been supported by the funding for the ByoPiC project from the European Research Council (ERC) under the European Union’s Horizon 2020 research and innovation programme grant agreement ERC-2015-AdG 695561.

%Klaus, Jenny, Nabila
KD acknowledges support by the COMPLEX project from the European Research Council (ERC) under the European Union’s Horizon 2020 research and innovation program grant agreement ERC-2019-AdG 882679 as well as support by the Deutsche Forschungsgemeinschaft (DFG, German Research Foundation) under Germany’s Excellence Strategy-EXC-2094-390783311. JS, KD, and NA acknowledge support by the grant agreements ANR-21-CE31-0019/490702358 from the French Agence Nationale de la Recherche / DFG for the LOCALIZATION project.

%Computing time
The authors acknowledge the Gauss Centre for Supercomputing e.V. (\url{www.gauss-centre.eu}) and GENCI (\url{https://www.genci.fr/}) for funding this project by providing computing time on the GCS Supercomputer SuperMUC-NG at Leibniz Supercomputing Centre (\url{www.lrz.de}) and Joliot-Curie at TGCC (\url{http://www-hpc.cea.fr}).

%SDSS credits
Funding for the SDSS and SDSS-II has been provided by the Alfred P. Sloan Foundation, the Participating Institutions, the National Science Foundation, the U.S. Department of Energy, the National Aeronautics and Space Administration, the Japanese Monbukagakusho, the Max Planck Society, and the Higher Education Funding Council for England. The SDSS Web Site is \url{http://www.sdss.org/}.

The SDSS is managed by the Astrophysical Research Consortium for the Participating Institutions. The Participating Institutions are the American Museum of Natural History, Astrophysical Institute Potsdam, University of Basel, University of Cambridge, Case Western Reserve University, University of Chicago, Drexel University, Fermilab, the Institute for Advanced Study, the Japan Participation Group, Johns Hopkins University, the Joint Institute for Nuclear Astrophysics, the Kavli Institute for Particle Astrophysics and Cosmology, the Korean Scientist Group, the Chinese Academy of Sciences (LAMOST), Los Alamos National Laboratory, the Max-Planck-Institute for Astronomy (MPIA), the Max-Planck-Institute for Astrophysics (MPA), New Mexico State University, Ohio State University, University of Pittsburgh, University of Portsmouth, Princeton University, the United States Naval Observatory, and the University of Washington.

%2MASS credits
This publication makes use of data products from the Two Micron All Sky Survey, which is a joint project of the University of Massachusetts and the Infrared Processing and Analysis Center/California Institute of Technology, funded by the National Aeronautics and Space Administration and the National Science Foundation.

%Cosmohub credits
This work has made use of CosmoHub. CosmoHub has been developed by the Port d'Informaci{\'o} Científica (PIC), maintained through a collaboration of the Institut de Física d'Altes Energies (IFAE) and the Centro de Investigaciones Energ{\'e}ticas, Medioambientales y Tecnol{\'o}gicas (CIEMAT) and the Institute of Space Sciences (CSIC \& IEEC), and was partially funded by the "Plan Estatal de Investigaci{\'o}n Científica y T{\'e}cnica y de Innovaci{\'o}n" program of the Spanish government.

%Cosmosim credits
The CosmoSim database used in this paper is a service by the Leibniz-Institute for Astrophysics Potsdam (AIP). The MultiDark database was developed in cooperation with the Spanish MultiDark Consolider Project CSD2009-00064. The authors gratefully acknowledge the Gauss Centre for Supercomputing e.V. (\url{www.gauss-centre.eu}) and the Partnership for Advanced Supercomputing in Europe (PRACE, \url{www.prace-ri.eu}) for funding the MultiDark simulation project by providing computing time on the GCS Supercomputer SuperMUC at Leibniz Supercomputing Centre (LRZ, \url{www.lrz.de}). The Bolshoi simulations have been performed within the Bolshoi project of the University of California High-Performance AstroComputing Center (UC-HiPACC) and were run at the NASA Ames Research Center.
\end{acknowledgements}

\bibliographystyle{aa}
\bibliography{simulation_coma}

\begin{appendix}
\section{Masses for SDSS galaxies}
\label{appendix_masses}
Galaxy masses were obtained from the Max Planck Institute for Astrophysics-Johns Hopkins University Value Added Catalogue (MPA-JHU VAC\footnote{\url{https://www.sdss.org/dr17/spectro/galaxy_mpajhu/}}). Spectral quantities in this catalogue were obtained with the \textsc{galSpec} software, following \citet{Brinchmann2004, Kauffmann2003, Tremonti2004}. Masses were derived with a fit to the $ugriz$ photometry corrected with spectral information, following \citet{Kauffmann2003, Salim2007}. We make use of the median value of the mass distribution output from the Bayesian fit to the photometry, available in the catalogue. We start with an initial sample of $927\,552$ galaxies in the MPA-JHU VAC, from which we remove duplicate observations and galaxies not in the MGS or in special releases (\textsc{primtarget} \& 64, \textsc{release} $\ne$ "extra", "special", and "extraspecial"). We select only objects classified as galaxies spectroscopically (\textsc{spectrotype} = "galaxy") and with good redshift measurements (\textsc{zwarning} = 0 and $z > 0$). The result of these selections is a sample of $654\,324$ galaxies which we match in position to the Legacy MGS with a tolerance of $0.5 \arcsec$. We check that the redshift of the matched objects is consistent between the MPA-JHU VAC and the Legacy MGS and we identify $520\,682$ galaxies with measured stellar mass. We stress that in this work we make use of galaxy masses only to calibrate our assumptions for the selection of halos corresponding to galaxies in the numerical simulation (see Section \ref{idgalinsim}) and we do not derive physical conclusions on the Coma system or its filaments based on this data set.

\begin{figure}
\centering
\resizebox{\hsize}{!}{\includegraphics{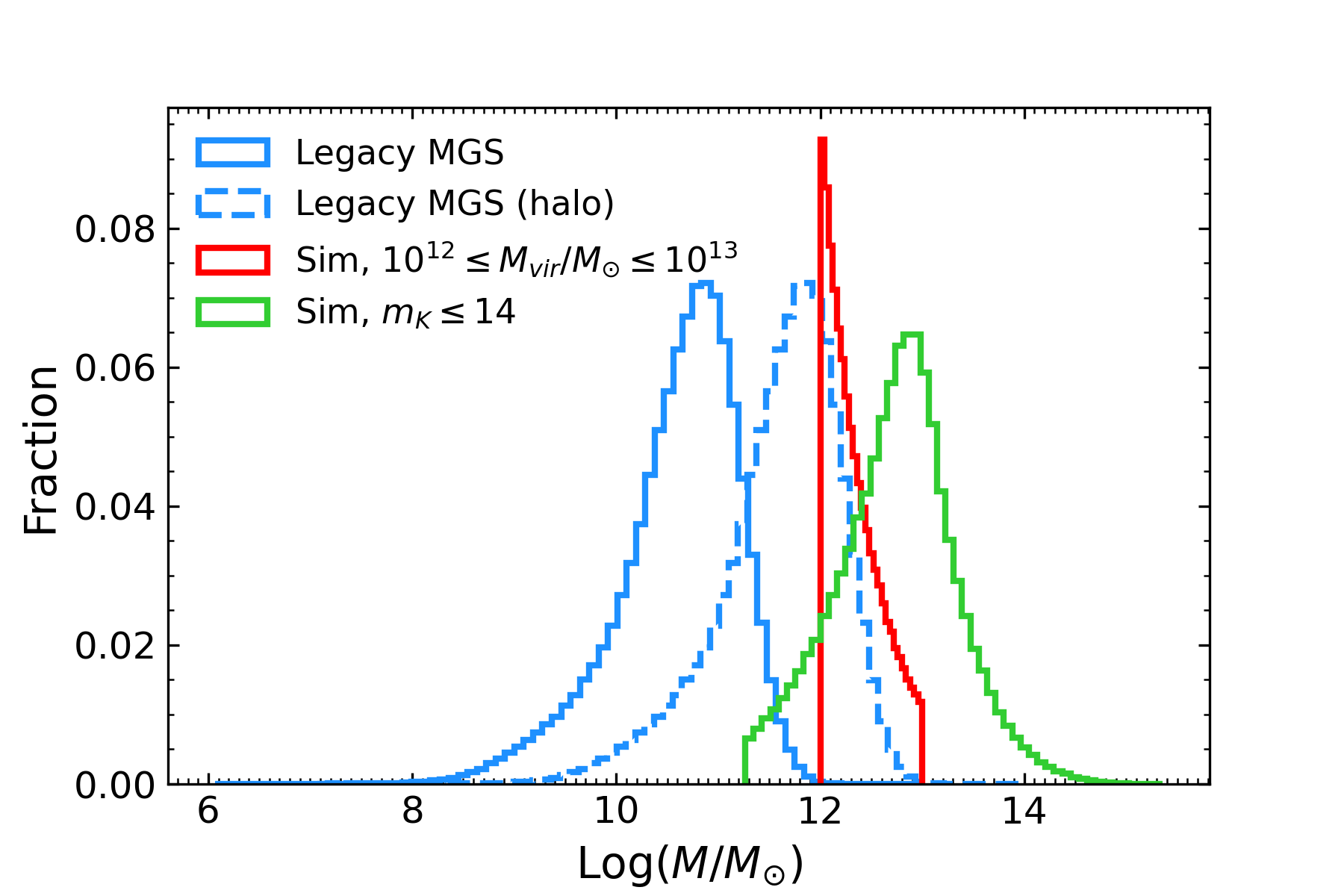}}
\caption{Mass distributions for the halos in our simulation and for SDSS galaxies. The solid light blue line shows the mass distribution for Legacy MGS galaxies, while the dashed light blue line is the same distribution shifted by one order of magnitude to provide a reference for the expected position of the mass distribution of the halos hosting the Legacy MGS galaxies. The red line is the mass distribution for simulated galaxies selected through a cut in mass and the green line is the mass distribution of simulated galaxies selected through a cut in magnitude. We note that the mass reported on the $x$-axis is the stellar mass for SDSS galaxies and the virial mass for the halos and galaxies in our simulation.}
\label{mdist}
\end{figure}

Figure \ref{mdist} shows the mass distribution for the halos and galaxies in our simulation and for the Legacy MGS galaxies (with mass derived by matching the Legacy MGS sample with the MPA-JHU VAC sample). The mass distribution of our mass-selected galaxy sample is largely consistent with the expected mass distribution for the halos hosting the galaxies of the Legacy MGS (this distribution was derived by shifting the mass distribution of Legacy MGS galaxies by one order of magnitude to higher masses, for reference). As for the mass distribution of our magnitude-selected simulated galaxies, it is somewhat overlapping with the expected mass distribution of the halos hosting the Legacy MGS galaxies, although marginally. However we stress that the matching of the Legacy MGS with the MPA-JHU VAC is itself somewhat uncertain, while the expected distribution of the halo mass of the Legacy MGS galaxies is only for reference. We therefore use this figure only to check our parameter selection rather than to drive it.

The expected mass distribution for the halos hosting the galaxies of the Legacy MGS was derived in this case assuming a Stellar-to-Halo Mass Relation (SHMR) expressed in terms of a fixed ratio of halo to stellar mass of 1 dex. We do note that this approach, while sufficient for us to check our parameter selection for the extraction of a simulated galaxy population from our parent halo population, may not reflect the true distribution of mass for halos hosting the Legacy MGS galaxies that could be obtained using variable SHMRs found in the literature. We have therefore compared the halo mass distribution of our mass and magnitude selected galaxies with the expected mass distributions for halos hosting Legacy MGS galaxies obtained using SHMRs from the literature.

In particular, we used Equation 6 (with best-fit values from the first row of Table 1) of \citet{Girelli2020} to derive the stellar mass distribution for halos in our mass-selected and magnitude-selected simulated galaxy samples. We find these to be in very good agreement with the stellar mass distribution for Legacy MGS galaxies obtained through the matching with the MPA-JHU VAC. 

We also used the Data Release 1 (DR1) of the \textsc{UniverseMachine} code applied to the Bolshoi-Planck simulation \citep{Klypin2016, RodriguezPuebla2016} by \citet{Behroozi2019}\footnote{\url{https://www.peterbehroozi.com/data.html}}. We make use of the light-cones created to reproduce the COSMOS field. While we are aware that this means a limited area and therefore a restricted number of halos, we can use this light-cone to select objects at $z \leq 0.3$ (to match the redshift limit of the Legacy MGS) and in two mass ranges: $10^{12} \leq M_{\mathrm{vir}}/M_{\sun} \leq 10^{13}$ (to simulate our mass-selected simulated galaxy sample) and $M_{\mathrm{vir}}/M_{\sun} \geq 10^{12}$ (to simulate the absence of an upper mass limit as is the case for our magnitude-selected simulated galaxy sample, although the actual mass distribution may differ). We then compare the stellar mass distribution for these halo samples, as derived by \citet{Behroozi2019}. We find good agreement with the stellar mass distribution for Legacy MGS galaxies obtained through the matching with the MPA-JHU VAC for both cases. 

Additionally, we use the simulated galaxies generated by \citet{Cora2018} by applying the semi-analytic model of galaxy formation \textsc{SAG} \citep{Cora2006, Lagos2008, Orsi2014, Gargiulo2015, MunozArancibia2015, Cora2018} to the Multidark MDPL2 simulation \citep{Klypin2016}\footnote{Downloaded from CosmoSim (\url{https://www.cosmosim.org}).}. We select halos with $M_{200\mathrm{c}} \geq 10^{12} M_{\sun}$ and $z \leq 0.3$. We then derived their total stellar mass by summing the stellar mass of the disk and spheroid components and we compared its distribution to the stellar mass distribution for Legacy MGS galaxies, finding good agreement.

Finally, we use the MICECATv1.0\footnote{Downloaded from CosmoHub (\url{https://cosmohub.pic.es/home} \citealt{Carretero2017, Tallada2020})} catalogue of \citet{Carretero2015}, created from the Marenostrum Institut de Ci{\`e}ncies de l’Espai (MICE) Grand Challenge (MICE-GC) simulation \citep{Crocce2015, Fosalba2015a, Fosalba2015b}. From this halo catalogue with photometric quantities computed by \citet{Carretero2015}, we selected halos with $z \leq 0.3$ and apparent $r$-band magnitude $r \leq 17.77$ to mimic the SDSS selection. We then compared the halo mass distribution to the halo mass distribution for mass-selected and magnitude-selected galaxies. We find close agreement.

Our conclusion is that although choosing a fixed SHMR may not be the most accurate way to convert between halo and stellar mass, it provides a good enough approximation for our goal of checking whether our parameter choice for the selection of a simulated galaxy sample is correct. We postpone to future works the use of a semi-analytical model to populate halos with galaxies in a realistic way.

\section{Extra projection}
\label{extraproj}
In this section we provide an additional projection to the one showed in Figure \ref{visual_skel_comparison_proj1}. This projection is located on the plane of the sky and shows the NE-W filament system in the vicinity of Coma. This figure shows filaments leaving the simulated Coma cluster more or less at the correct directions.

\begin{figure*}
\sidecaption
\includegraphics[trim = 0.75cm 4cm 1cm 4.2cm, clip = true, width = 12cm]{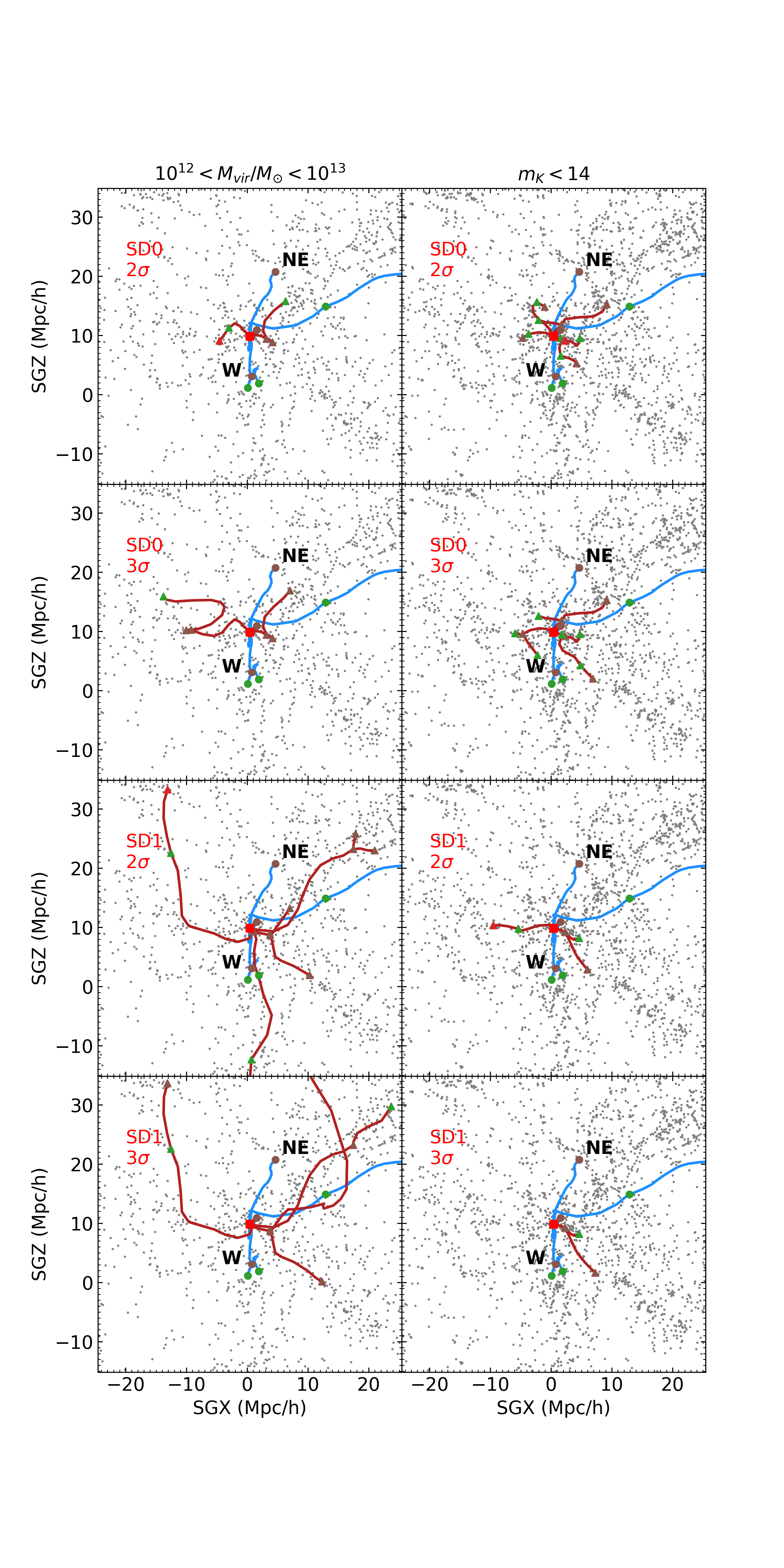}
\caption{Filaments in the constrained simulation compared to filaments in observations. The left column shows the case of a mass selected galaxy sample in the constrained simulation, the right column shows the case of a magnitude selected galaxy sample. In each panel, simulated galaxies are shown as grey points. Blue lines are filaments from \citet{Malavasi2020Coma} in supergalactic coordinates, the red square marks the position of the Coma cluster. Dark red lines are filaments obtained in the simulation with a variety of \disperse parameter combinations, namely: $\mathrm{SD0} - 2\sigma$ (top row), $\mathrm{SD0} - 3\sigma$ (second row), $\mathrm{SD1} - 2\sigma$ (third row), $\mathrm{SD1} - 3\sigma$ (bottom row). Circle and triangles mark the positions of critical points (in observations and simulations, respectively) and are color-coded according to their type (red: maxima, green: type 2 saddles, brown: bifurcations). This figure shows the SGX-SGZ projection (in this projection, the increasing redshift direction is along the y axis). The observed NE and W filaments are marked for reference.}
\label{visual_skel_comparison_proj3}
\end{figure*}
\end{appendix}

\end{document}